\documentclass{aa}
\usepackage{graphicx}
\usepackage{txfonts}
\usepackage{hyperref}
\usepackage{color}
\usepackage{amsmath}
\usepackage{amssymb}
\usepackage{xspace}

\providecommand{\sorthelp}[1]{}

\newcommand{\emm}[1]{\ensuremath{#1}}
\newcommand{\emr}[1]{\emm{\mathrm{#1}}}
 
\newcommand{\unit}[1]{\emr{\,#1}}
\newcommand{\eq}[1]{\begin{equation}#1\end{equation}}
\newcommand{\bA}{\ensuremath{\boldsymbol{A}}\xspace}
\newcommand{\bI}{\ensuremath{\boldsymbol{I}}\xspace}
\newcommand{\bchi}{\ensuremath{\boldsymbol{\chi}}\xspace}
\newcommand{\btheta}{\ensuremath{\boldsymbol{\theta}}\xspace}
\newcommand{\var}{\mathrm{var}}
\newcommand{\bias}{\mathrm{bias}}
\newcommand{\CRB}{{\cal{B}}}
\newcommand{\dco}{\ensuremath{^{12}\mathrm{CO}}\xspace}
\newcommand{\tco}{\ensuremath{^{13}\mathrm{CO}}\xspace}
\newcommand{\cdo}{\ensuremath{\mathrm{C}^{18}\mathrm{O}}\xspace}
\newcommand{\dcouz}{\ensuremath{^{12}\mathrm{CO}}(1\emm{-}0)\xspace}
\newcommand{\tcouz}{\ensuremath{^{13}\mathrm{CO}}(1\emm{-}0)\xspace}
\newcommand{\cdouz}{\ensuremath{\mathrm{C}^{18}\mathrm{O}}(1\emm{-}0)\xspace}
\newcommand{\dcodu}{\ensuremath{^{12}\mathrm{CO}}(2\emm{-}1)\xspace}
\newcommand{\tcodu}{\ensuremath{^{13}\mathrm{CO}}(2\emm{-}1)\xspace}
\newcommand{\cdodu}{\ensuremath{\mathrm{C}^{18}\mathrm{O}}(2\emm{-}1)\xspace}

\newcommand{\dcoqt}{\ensuremath{^{12}\mathrm{CO}}(4\emm{-}3)\xspace}
\newcommand{\dcocq}{\ensuremath{^{12}\mathrm{CO}}(5\emm{-}4)\xspace}
\newcommand{\dcosc}{\ensuremath{^{12}\mathrm{CO}}(6\emm{-}5)\xspace}
\newcommand{\uz}{1\emm{-}0\xspace}
\newcommand{\du}{2\emm{-}1\xspace}
\newcommand{\radec}[6]{\emr{#1^{h}#2^{m}#3^{s},#4^{\circ}#5^{'}#6^{''}}}

\begin{document} 

\title{\cdo, \tco, and \dco abundances \\
  and excitation temperatures in the Orion B molecular cloud}

\subtitle{An analysis of the precision achievable when modeling spectral
  line \\
  within the Local Thermodynamic Equilibrium approximation}

\author{%
  Antoine Roueff\inst{\ref{Marseille}} %
  \and Maryvonne Gerin\inst{\ref{LERMA}} %
  \and Pierre Gratier \inst{\ref{LAB}} %
  \and François Levrier\inst{\ref{LPENS}} %
  \and Jérôme Pety\inst{\ref{IRAM},\ref{LERMA}} %
  \and Mathilde Gaudel\inst{\ref{LERMA}} %
  \and Javier R. Goicoechea\inst{\ref{CSIC}} %
  \and Jan H. Orkisz\inst{\ref{Chalmers}} %
  \and Victor de Souza Magalhaes\inst{\ref{IRAM}} %
  \and Maxime Vono\inst{\ref{IRIT}} %
  \and S\'ebastien Bardeau\inst{\ref{IRAM}} %
  \and Emeric Bron\inst{\ref{Meudon}} %
  \and Jocelyn Chanussot\inst{\ref{Grenoble}} %
  \and Pierre Chainais\inst{\ref{Lille}} %
  \and Viviana V. Guzman\inst{\ref{Catholica}} %
  \and Annie Hughes\inst{\ref{IRAP}} %
  \and Jouni Kainulainen\inst{\ref{Chalmers}} %
  \and David Languignon\inst{\ref{Meudon}} %
  \and Jacques Le Bourlot\inst{\ref{Meudon}} %
  \and Franck Le Petit\inst{\ref{Meudon}} %
  \and Harvey S. Liszt\inst{\ref{NRAO}} %
  \and Antoine Marchal\inst{\ref{CITA}} %
  \and Marc-Antoine Miville-Deschênes\inst{\ref{CEA}} %
  \and Nicolas Peretto\inst{\ref{UC}} %
  \and Evelyne Roueff\inst{\ref{Meudon}} %
  \and Albrecht Sievers\inst{\ref{IRAM}}}

\institute{%
  Aix Marseille Univ, CNRS, Centrale Marseille, Institut Fresnel, Marseille, France,  \email{antoine.roueff@fresnel.fr}. \label{Marseille} %
  \and LERMA, Observatoire de Paris, PSL Research University, CNRS, Sorbonne Universit\'es, 75014 Paris, France. \label{LERMA} %
  \and Laboratoire d'Astrophysique de Bordeaux, Univ. Bordeaux, CNRS,  B18N, Allee Geoffroy Saint-Hilaire,33615 Pessac, France. \label{LAB} %
  \and Laboratoire de Physique de l’Ecole normale supérieure, ENS,
  Université PSL, CNRS, Sorbonne Université, Université de Paris, Sorbonne Paris Cité, Paris, France. \label{LPENS} %
  \and IRAM, 300 rue de la Piscine, 38406 Saint Martin d'H\`eres,  France. \label{IRAM} %
  \and Instituto de Física Fundamental (CSIC). Calle Serrano 121, 28006, Madrid, Spain. \label{CSIC} %
  \and Chalmers University of Technology, Department of Space, Earth and Environment, 412 93 Gothenburg, Sweden. \label{Chalmers} %
  \and University of Toulouse, IRIT/INP-ENSEEIHT, CNRS, 2 rue Charles Camichel, BP 7122, 31071 Toulouse cedex 7, France. \label{IRIT} %
  \and LERMA, Observatoire de Paris, PSL Research University, CNRS, Sorbonne Universit\'es, 92190 Meudon, France. \label{Meudon} %
  \and Univ. Grenoble Alpes, Inria, CNRS, Grenoble INP, GIPSA-Lab,
  Grenoble, 38000, France. \label{Grenoble} %
  \and Univ. Lille, CNRS, Centrale Lille, UMR 9189 - CRIStAL, 59651 Villeneuve d’Ascq, France. \label{Lille} %
  \and Instituto de Astrofísica, Pontificia Universidad Católica de Chile, Av. Vicuña Mackenna 4860, 7820436 Macul, Santiago, Chile. \label{Catholica} %
  \and Institut de Recherche en Astrophysique et Planétologie (IRAP), Université Paul Sabatier, Toulouse cedex 4, France. \label{IRAP} %
  \and National Radio Astronomy Observatory, 520 Edgemont Road, Charlottesville, VA, 22903, USA. \label{NRAO} %
  \and Canadian  Institute  for  Theoretical  Astrophysics,  Universityof  Toronto,  60  Saint  George  Street,  14th  floor,  Toronto,  ON,M5S 3H8, Canada. \label{CITA} %
  \and AIM, CEA, CNRS, Université Paris-Saclay, Université Paris Diderot,
  Sorbonne Paris Cité, 91191 Gif-sur-Yvette, France. \label{CEA}%
  \and School of Physics and Astronomy, Cardiff University, Queen's buildings, Cardiff CF24 3AA, UK. \label{UC} %
} 

\date{} 


\abstract
{CO isotopologue transitions are routinely observed in molecular clouds to
  probe the column density of the gas, the elemental ratios of carbon and
  oxygen, and to trace the kinematics of the environment.}
{We aim at estimating the abundances, excitation temperatures, velocity
  field and velocity dispersions of the three main CO isotopologues towards
  a subset of the Orion B molecular cloud, which includes IC\,434,
  NGC\,2023, and the Horsehead pillar.}
{We use the Cramer Rao Bound (CRB) technique to analyze and estimate the
  precision of the physical parameters in the framework of
  local-thermodynamic-equilibrium (LTE) excitation and radiative transfer
  with an additive white Gaussian noise.
  We propose a maximum likelihood estimator to infer the physical
  conditions from the \uz{} and \du{} transitions of CO isotopologues.
  Simulations show that this estimator is unbiased and efficient for a
  common range of excitation temperatures and column densities
  ($T_\emr{ex}> 6\unit{K}$, $N> 10^{14}\,-\,10^{15}~\unit{cm^{-2}}$).  }
{Contrary to the general assumptions, the different CO isotopologues have
  distinct excitation temperatures, and the line intensity ratios between
  different isotopologues do not accurately reflect the column density
  ratios. We find mean fractional abundances that are consistent with
  previous determinations towards other molecular clouds. However,
  significant local deviations are inferred, not only in regions exposed to
  UV radiation field but also in shielded regions. These deviations result
  from the competition between selective photodissociation, chemical
  fractionation, and depletion on grain surfaces. We observe that the
  velocity dispersion of the C$^{18}$O emission is 10\% smaller than that
  of $^{13}$CO. The substantial gain resulting from the simultaneous
  analysis of two different rotational transitions of the same species is
  rigorously quantified.}
{The CRB technique is a promising avenue for analyzing the estimation of
  physical parameters from the fit of spectral lines. Future work will
  generalize its application to non-LTE excitation and radiative transfer
  methods.}



\maketitle{} 

\section{Introduction}

Spectroscopic measurements are commonly used to probe astrophysical
objects. In the interstellar medium, the moderate temperatures and
densities of diffuse and molecular clouds
($T_\emr{kin} \sim 10 - 100\unit{K}$, and
$n \sim 10^2 - 10^5\emr{cm}^{-3}$, \citealt{draine:11}) are well suited for
the emission in the low energy rotational lines of molecules such as carbon
monoxide, which are accessible at millimeter wavelengths. The advent of
sensitive broadband heterodyne receivers provides homogeneous data sets of
various CO isotopologues and other species with high signal-to-noise ratios
over large fields of view. The ORION-B IRAM-30m large program (Outstanding
Radio-Imaging of OrioN B, co-PIs: J.~Pety and M.~Gerin) aims at imaging 5
square degrees towards the southern part of the Orion B molecular cloud
over most of the 3\,mm atmospheric window. Carbon monoxide is especially
interesting because it is one of the most abundant molecules after
molecular hydrogen. Using the unsupervised meanshift clustering method on
the intensities of the CO isotopologues, \citet{bron:18} show that it is
possible to cluster the emission line data across the analyzed field of
view into a few classes of increasing (column) densities. In two empirical
studies, \citet{gratier:17} and~Gratier et al. (submitted as a companion
paper) show qualitatively and quantitatively that the \dcouz, \tcouz, and
\cdouz lines are indeed tracing the molecular gas well. Their quantitative
comparisons show that the \emr{H_2} column density deduced from the dust
emission can be accurately estimated from the \dcouz, \tcouz, and \cdouz
lines in the column density range from $10^{21}$ to
$\gtrsim 10^{22}\unit{cm^{-2}}$.

Assuming identical excitation temperatures, the opacity of the ground state
transitions is expected to be smaller for \tco than for \dco, and even
smaller for \cdo because of the difference in elemental abundances,
\emr{^{12}C/^{13}C \sim 60} and \emr{^{16}O/^{18}O \sim
  500}~\citep{langer:90,wilson:94}.  These three lines can thus be used to
probe progressively higher gas column densities, provided the relative
elemental abundances are constant and the CO isotopologue abundances track
the elemental abundances. However chemical models and observations show
that selective photodissociation and carbon isotopic fractionation can
significantly modify the relative abundances of carbon monoxide
isotopologues, as compared to elemental
abundances~\citep{visser:09,liszt:17,roueff:15}. Fractionation via the
exchange reaction between $^{13}$C$^+$ and $^{12}$CO leads to an
enhancement of the $^{13}$CO abundance in the diffuse/translucent regions
where CO and C$^+$ coexist and the kinetic temperature remains moderate
($\lesssim 50\unit{K}$, \citealt{liszt:12}). This mechanism widens the \tco
emitting region and brings it closer to that of \dco, which favors the
simultaneous detection of both isotopologues on wide fields of
view. However the ratio of isotopologue abundances can be significantly
different from the ratio of elemental abundances, which complicates the
determination of the \emr{^{13}C} elemental abundance from CO observations
only.  Up to now, the most reliable determinations of the
\emr{^{12}C/^{13}C} elemental abundance ratio have been obtained using
\emr{C^+} or C observations in regions without significant fractionation
\citep[e.g.][]{keene:98,ossenkopf:13}, or involve \cdo and the doubly
isotopic species $^{13}$C$^{18}$O \citep{langer:90}.

No such fractionation reaction exists for oxygen. However, the more
abundant CO isotopologues shield themselves from the destructive effect of
UV photons more efficiently than less abundant isotopologues because the
photodissociation of carbon monoxide is governed by line absorption.  This
effect called selective photodissociation is important. It has been studied
in detail through laboratory experiments \citep[e.g.][]{stark:14} and in
models of photo-dissociation regions \citep[e.g.][]{visser:09}.  In
observations, it is clearly seen as an offset between the threshold for the
apparition of \dco (near $A_\emr{V} = 0.5\,\emr{mag}$) and \cdo{}
$(1.5\,\emr{mag})$ in the Taurus molecular cloud, and this offset is not
due to a difference in the detection
sensitivity~\citep{frerking:89,cernicharo:87}.  Typically, the $^{13}$CO
abundance is enhanced through fractionation in the same regions where the
C$^{18}$O abundance decreases due to selective photodissociation. This
leads to a broad range of the $^{13}$CO/C$^{18}$O abundance ratio for a
given set of elemental abundances.

Determining the ratio of elemental abundances of the C and O isotopes is
interesting because it provides information on the stellar populations
which have produced these elements.  Some external galaxies exhibit CO
isotopologue ratios that significantly differ from the expected value based
on the mean elemental abundances in the solar neighborhood. Such
differences can trace differences in elemental abundances, hence in stellar
populations and IMF shape~\citep{sliwa:17,martin:19}.  However, a proper
account of isotopic chemistry described above must be performed in order to
use the information on the relative abundances of the CO isotopologues.

Finally, \citet{orkisz:19} show, in an analysis of the filamentary
structure of the Orion B molecular cloud, that the gas velocity dispersion
determined from \cdo{} reaches a minimum value in the filament ridges, that
it is always lower than the velocity dispersion determined by \tco{}. This
suggests that this variation of velocity dispersion between CO isotopologue
traces the dissipation of turbulence when entering the dense filaments
inside molecular clouds.

Constraining all these astrophysical effects relies on a precise derivation
of physical conditions and chemical composition from spectroscopic
observations.  This in turn relies on the resolution of the radiative
transfer equation because the line intensities and profiles bear
information on the line emission mechanisms. The large data volumes
provided by observational programs like ORION-B require new statistical
analysis methods using the information in an optimal way, and a derivation
of the physical parameters and their associated errors with a rigorous
methodology. For instance, the emission of the lowest rotational
transitions of the three major isotopologues of carbon monoxide, $^{12}$CO,
$^{13}$CO, and C$^{18}$O is commonly used to determine the molecular gas
column density and evaluate the mass of molecular gas. Because these lines
can now be observed simultaneously, leading to an homogeneous flux
calibration and therefore precise relative calibration, it is essential to
have a good estimation of the precision on the mass estimate.

In estimation theory, the Cramer Rao bound (CRB) provides a precision of
reference that does not depend on a specific estimator of the searched
quantity, but only on the physical model and the statistical properties of
the noise~\citep[see, e.g.,][]{bonaca:18,espinosa:18}. The CRB further
allows the quantification of the loss of precision due to degeneracies
between the estimated parameters (for instance column density and
excitation temperature). Hence, a large value of this bound indicates
insufficient data or knowledge with respect to a given physical model. We
will here apply this technique in the simplest possible model framework,
i.e., the emission of lines in Local Thermodynamic Equilibrium (LTE), which
can be fully expressed with analytical equations.

The level populations of interstellar molecules result from the balance of
collisional (and possibly radiative) excitation and radiative \&
collisional de-excitation.  Therefore the level populations often deviate
from LTE conditions because the collisions are not efficient enough to
populate all energy levels according to a Boltzmann distribution. With its
low dipole moment (0.1 Debye) and high abundance relative to H$_2$, the low
energy rotational lines of carbon monoxide are bright and easily
thermalized in collisions with H$_2$, H and He. This means that the LTE
model is still a good approximation for this molecule, i.e., the rotational
level populations can be described by a Boltzman distribution at a single
excitation temperature \citep{liszt:06,liszt:76,goldsmith:99,goldreich:74}.
Deviations from the LTE model have been theoretically studied. For
instance, using non local, non LTE radiative transfer models of a uniform
(constant density and temperature) spherical cloud, \citet{bernes:79} shows
that the excitation temperatures of the \dcouz and \dcodu lines exhibit
moderate spatial variations from edge to center.  It is concluded that the
LTE model is mostly valid for the ground state transition and deviations
from this approximation increase with the quantum number of the upper level
\citep{vandertak:07}.

With a wide range of physical conditions, from bright far-UV illuminated
regions to cold and shielded regions through diffuse and translucent gas
irradiated by a moderate radiation field, the Orion B molecular cloud is an
ideal place to probe to which extent fractionation and selective
photodissociation can modify the elemental abundance ratio. It is also a
good region to probe the differences in excitation between isotopologues as
the simple hypothesis of equal excitation temperatures for \dco, \tco and
\cdo{} may not be valid, as discussed in~\citet{bron:18}.

The article is organized as follows. Section~\ref{sec_data} presents the
data used in this paper. Section~\ref{sec_model} summarizes the
mathematical formulation of the LTE radiative
transfer. Section~\ref{sec_CRB} computes and analyzes the precision
achievable for this theoretical framework. Section~\ref{sec_real_data}
illustrates the proposed methodology on actual data
sets. Section~\ref{sec_astro} focuses on the astrophysical interpretations
of these results. Appendix~\ref{sec_Fisher} details the calculation of some
gradients necessary to compute the Fisher
matrix. Appendix~\ref{sec_MLE_definition} describes our implementation of
the maximum likelihood estimator, and appendix~\ref{sec_MLE_performance}
discusses the performance of this estimator.


\newcommand{\FigSignalMaps}{%
  \begin{figure*}
    \centering{} %
    \includegraphics{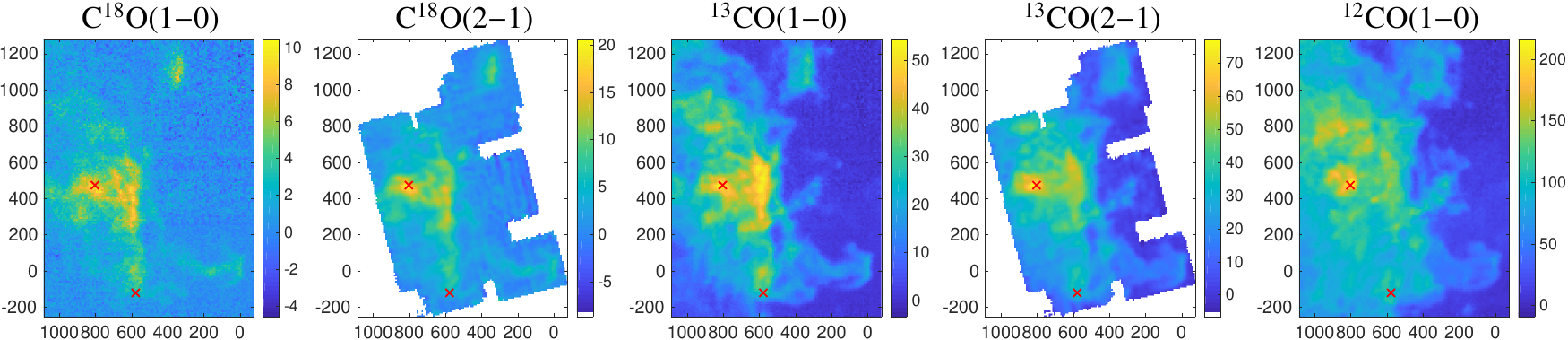}
    \caption{Spatial distributions of the integrated intensity (in
      \unit{K\,km\,s^{-1}}) of the considered lines.  The maps have been
      rotated counterclockwise by 14 degrees from the RA/DEC J2000
      reference frame. The spatial offsets are given in arcsecond from the
      projection center located at
      \radec{05}{40}{54.270}{-02}{28}{00.00}. Red crosses stand for two
      particular lines of sight which are analyzed in
      Figure~\ref{Fig_real_data_spectres}.}
    \label{Fig_Ima_in_pixels}
  \end{figure*}
}

\newcommand{\TabLines}{%
  \begin{table}
    \centering %
    \caption{Properties of observed lines.}
    \begin{tabular}{cccccr}
      \hline
      \hline
      Species   & Line  & $\nu$      & $dV$\tablefootmark{1} & Beam\tablefootmark{2} & Noise\tablefootmark{3} \\
                &       & MHz        & \unit{km\,s^{-1}}     & $''$                  & mK    \\
      \hline
      C$^{18}$O & 1$-$0 & 109782.173 & 0.5               & 23.5 & 116\\
      C$^{18}$O & 2$-$1 & 219560.319 & 0.5               & 23.5 &  96\\
      $^{13}$CO & 1$-$0 & 110201.354 & 0.5               & 23.5 & 116\\
      $^{13}$CO & 2$-$1 & 220398.686 & 0.5               & 23.5 & 134\\
      $^{12}$CO & 1$-$0 & 115271.202 & 0.5               & 23.5 & 278\\
      \hline
    \end{tabular}
    \tablefoot{%
      \tablefoottext{1}{Channel spacing after resampling.} %
      \tablefoottext{2}{Angular resolution after smoothing.} %
      \tablefoottext{3}{Median noise $\sigma_b$ after resampling and
        smoothing.}}
    \label{tab:lines}
  \end{table}}


\FigSignalMaps{}%
\TabLines{}%

\section{Description of the data}
\label{sec_data}

We will try to estimate the velocity field, the column density, and the
excitation temperature of the CO isotopologues from the analysis of the
\tcouz, \tcodu, \cdouz, \cdodu and \dcouz lines towards parts of the Orion
B molecular cloud. We will compare our results with the dust-traced H$_2$
column density and dust temperature. This section describes the associated
data sets.

\subsection{IRAM-30m observations}

\subsubsection{3\,mm CO lines from the ORION-B large program}

The 3\,mm data were obtained with the IRAM-30m as part of the ORION-B large
program. \citet{pety:17} present in detail the acquisition and reduction of
the dataset used in this study. In short, the used data were acquired at
the IRAM-30m telescope using the EMIR receiver and Fourier transform
spectrometer from August 2013 to November 2014. The frequency range from 84
to 116\,GHz was completely sampled at 200\,kHz spectral resolution.  The
$J=1-0$ lines of the CO isotopologues analyzed here are observed in a
single receiver tuning. These lines are thus well inter-calibrated. The
absolute flux calibration at 3\,mm for the IRAM-30m telescope is estimated
to be better than 5\%.

\subsubsection{1\,mm CO lines}


The \tcodu and \cdodu were also observed at the IRAM-30m in 2006 (PI:
N.~Peretto) using the ABCD generation of receivers and the VESPA
auto-correlator. The two lines were observed simultaneously ensuring an
excellent inter-calibration.

Data reduction was carried out using the \texttt{GILDAS}\footnote{See
  \url{http://www.iram.fr/IRAMFR/GILDAS} for more information about the
  GILDAS software~\citep{pety:05}.}\texttt{/CLASS} software. The
contribution of the atmosphere was first removed (ON-OFF procedure) and the
data were calibrated to the $T_A^\star$ scale using the standard
chopper-wheel method~\citep{penzias:73}. The data were then converted to
main-beam temperatures using the standard forward (0.94) and main-beam
(0.62) efficiencies for the ABCD receiver around 220\,GHz\footnote{For
  details, see
  \url{http://www.iram.es/IRAMES/mainWiki/Iram30mEfficiencies}.}. The
resulting absolute flux calibration is estimated to be better than 10\%. We
subtracted a first order baseline from every spectrum, excluding the
velocity range from 5 to 15\unit{km\,s^{-1}} in the Local Standard of Rest
(LST) frame. The spectra were finally gridded into a data cube through a
convolution with a Gaussian kernel of full width at half maximum $\sim1/3$
of the IRAM-30m telescope beamwidth at the line rest frequency.

\subsection{Herschel observations}
\label{sec_herschel}

In order to get independent constraints on the physical conditions in the
Orion B cloud, we use the dust continuum observations from the
\textit{Herschel} Gould Belt Survey~\citep{andre:10,schneider:13} and from
the \textit{Planck} satellite~\citep{planck2011-1.1}. The fit of the
spectral energy distribution by~\citet{lombardi:14} gives us access to the
spatial distributions of the dust opacity at $850\unit{\mu m}$ and of the
dust temperature.  As in \citet{pety:17}, we converted
$\tau_{850\unit{\mu m}}$ to visual extinctions using
$A_\mathrm{V} = 2.7\times 10^4 \, \tau_{850} \, \mathrm{mag}$, and the
visual extinction into H$_2$ column density using
$N(\emr{H_2})/A_\emr{V} = 0.9 \times 10^{21}\,\emr{H\,cm^{-2}\,mag^{-1}}$.

\subsection{Field of view}

We will jointly analyze the $J=1-0$ and $J=2-1$ lines of the CO
isotopologues. We thus restrict the field of view to the region that was
observed at 3 and 1\,mm.  This covers $19' \times 26'$ towards the Orion B
molecular cloud part that contains the Horsehead nebula, and the
H\textsc{ii} regions NGC\,2023 and IC\,434. The cubes used here are rotated
counterclockwise by $14\degr$ around the projection center
(\radec{05}{40}{54.270}{-02}{28}{00.00}) in the RA/DEC J2000 reference
frame (see Fig.~\ref{Fig_Ima_in_pixels}). The coordinates are given in
offsets $(\delta x, \delta y)$ in arcseconds from this projection center.
The IRAM-30m angular resolution ranges from $11.5''$ at 220\,GHz to
$23.5''$ at 110\,GHz.  The position-position-velocity cubes of each line
were smoothed to a common angular resolution of $23.5''$ to avoid
resolution effects during the comparison. At a distance of
400\,pc~\citep{menten:07}, the sampled linear scales range from
$\sim0.045\unit{pc}$ to $\sim3\unit{pc}$.

The spectral and spatial axes were resampled in order to share the same
spatial grid and velocity axis for all lines. The spectroscopic
observations thus provide position-position-velocity cubes\footnote{The
  data products associated with this paper are available at
  \url{http://www.iram.fr/~pety/ORION-B}} of $129 \times 170 \times 80$
pixels, each pixel covering $9'' \times 9'' \times 0.5\unit{km\,s^{-1}}$
(Nyquist sampling at 3\,mm).  Figure~\ref{Fig_Ima_in_pixels} shows the maps
of the intensity integrated between 0 and $20\unit{km\,s^{-1}}$ for the
five lines of interest.

\subsection{Noise}

In this paper, the standard deviation of the noise $\sigma_{b}$ is
estimated only on negative values of each spectrum through
\begin{equation}
  \sigma_b=\left(\frac{1}{K_\emr{neg}}\sum_{k\in \{T_k \le 0\}}
    T_k^2\right)^{1/2},
\end{equation}
where $T$ is the intensity in Kelvin, and $K_\emr{neg}$ the number of
channels that have a negative value of the intensity. This allows us to
compute it without \textit{a priori} information on the velocity range
where the line appears, but it assumes that the baselining removed any
intensity offset. Table~\ref{tab:lines} lists the median noise estimated
after spectral resampling and angular smoothing.

\subsection{Line profiles}

A fraction of the studied field of view shows spectra that can only be
modeled with more than one velocity component along the line of
sight. While we will adapt our formalism to handle such cases, it is not
obvious to devise a robust statistical test to deduce the best number of
components that must be used. This is particularly true at transitions
between regions where the number of required velocity components changes to
get a good fit. To address this issue, we used the ROHSA \citep{marchal:19}
algorithm that makes a Gaussian decomposition based on a multi-resolution
process from coarse to fine grid. We only used here the spectra denoised by
ROHSA to provide a spatially coherent estimation of the number of
components and some initial estimation of their associated central
velocities for each pixel.



\section{Radiative Transfer in Local Thermodynamic Equilibrium}
\label{sec_model}

Molecular line emission and absorption in the case of Local Thermodynamic
Equilibrium (LTE) are well known~\citep[see, e.g.,][]{mangum:15}. In this
Section, we just summarize the associated notations and equations so that
we can easily explain the precision analysis framework on this case in the
next section. For the sake of simplicity, we focus on a single chemical
species and a single velocity component along one line of sight. The
observed spectrum as a function of frequency $\nu$ is defined as
\eq{ x(\nu)=s(\nu) + b(\nu),
  \label{eq_x} }
where $b$ is a (thermal) Gaussian noise, and $s$ is the spectrum associated
to the species of interest. The specific intensity $s$ and the associated
measurement noise $b$ are expressed in Kelvins following the standard use
in radioastronomy. The data reduction (atmospheric ON-OFF calibration and
spectrum baselining to subtract the slowly-varying continuum residual from
the receiver and the atmosphere) delivers a noise $b$ that is centered
(i.e., with zero-mean), and whose variance can be considered constant over
each line profile.

We assume that two lines $(l\in\{1,2\})$ from the same species are
observed. The photons of each line are emitted at the rest frequency of the
line, $\nu_{l}$, and redshifted in frequency because of the Doppler shift
due to the motion of the gas along the line of sight in the observation
frame, typically the Local Standard of Rest (LSR). The photon is thus
received at the redshifted frequency
$\nu^\emr{red}_{l}=\nu_{l}\left(1-\frac{V}{c}\right)$, where $V$ is the
velocity of the emitting cell of gas in the LSR frame and $c$ is the speed
of light. This equation is the radio low-velocity approximation of the
Doppler effect. The Doppler effect due to the motion of the observer
relative to the LSR is automatically taken into account in the data
acquisition process. Therefore, each line of the dataset is analyzed in the
LSR frame. In this frame each line is centered around a typical velocity,
noted $\Delta_V$. This velocity is related to the redshifted centroid
frequency of the line, $\nu^\emr{cent}_{l}$, through a particular case of
the previous equation
\eq{ \nu^\emr{cent}_{l}=\nu_{l}\left(1-\frac{\Delta_V}{c}\right).  }
We assume that the only background source of emission is the Cosmic
Microwave Background (CMB). In this case, the intensity $s$ at observed
frequency $\nu$ around $\nu^\emr{red}_{l}$ can be written as
\eq{ s(\nu)=\left\{ J(T_\emr{ex},\nu_{l})-J(T_\emr{CMB},\nu)\right\}
\left[1-\exp(-\Psi(\nu))\right]
  \label{eq_s_m} }
where $T_\emr{CMB}$ is the known CMB temperature
($T_\emr{CMB}=2.73\unit{K}$, \citealt{mather:94}), $T_\emr{ex}$ is the
unknown excitation temperature along the line of sight, and $J$ is a
measure of intensity at a given temperature
\eq{ J(T,\nu) = \frac{c^2}{2k\nu^2} B(T,\nu) = \frac{h \nu}{k}
\frac{1}{\exp{\frac{h \nu}{k T}} - 1},
  \label{eq_J} }
where $B(T,\nu)$ is the spectral distribution of the radiation of a black
body at temperature $T$. The term $\left[1-\exp(-\Psi(\nu))\right]$ in
Eq.~\eqref{eq_s_m} represents the emission/absorption by the
emitting/absorbing medium along the line of sight, considered as a uniform
slab. The function $\Psi$ is the profile that corresponds to the integrated
opacity through the whole slab. For each line $l$, it can be written as
\eq{ \Psi_l(\nu) = \alpha_{l}\, \phi\left(\nu;\nu^\emr{cent}_{l}, \nu_{l}
\frac{\sigma_{V}}{c}\right).
  \label{eq_tau_m} }
In this equation, $\sigma_{V}$ is the velocity dispersion of the source
along the line of sight. It varies as a function of the local physical
conditions (higher temperatures and higher turbulence will lead to larger
values). The function $\phi$ is a Gaussian profile
\eq{ \phi(\nu;\nu_o,\sigma_\nu)=\frac{1}{\sqrt{2\pi}\sigma_\nu}
\exp\left(-\frac{(\nu-\nu_o)^2}{2\sigma_\nu^2}\right),
  \label{eq_phi} }
where $\sigma_\nu$ is the frequency dispersion in the source rest frame. It
is related to $\sigma_{V}$ by $\sigma_\nu = \nu_{l}\,\sigma_{V}/c$, because
of the Doppler effect.  Finally, the amplitude $\alpha_{l}$ associated to
the Gaussian profile $\phi$ and line $l$ is
\eq{%
  \alpha_l= \frac{c^2}{8\pi}\frac{N}{Q(T_\emr{ex})}
  \frac{A_l \, g_\emr{up}}{ \nu_l^2}
  \exp\left[-\frac{E_\emr{up}}{T_\emr{ex}}\right]
  \left(\exp\left[\frac{h \,\nu_l}{k\,T_\emr{ex}}\right] - 1\right)
  \label{eq_alpha} }
where $A_l$ is the Einstein spontaneous emission rate for line $l$,
$g_\emr{up}$ is the degeneracy of the upper level of the line, $E_\emr{up}$
its energy (in units of Kelvin), and $N$ the column density of the species
along the line of sight.  The partition function $Q(T_\emr{ex})$ is
tabulated in molecular databases (e.g., CDMS, \citealt{muller:01}, or JPL,
\citealt{pickett:98}), and its temperature dependence can be interpolated
for each species.  The partition function is computed as the sum of the
populations of all energy levels $E_k$.  If the energy levels are expressed
in Kelvin, $Q(T_\emr{ex})$ can be written as
\eq{ Q(T_\emr{ex}) = \sum_{k=1}^{+\infty} g_k
\exp\left[-\frac{E_{k}}{T_\emr{ex}} \right].
  \label{eq:part} }
The parameter $\alpha_{l}$ is related to the line opacity
$\tau_l$ \eq{ \tau_{l}=\frac{\alpha _{l}\, c}{\sqrt{2\pi} \nu_l\sigma_{V}},
  \label{eq_tau_ml} }
which is dimensionless.
The excitation temperature is defined from the ratio of the population in
the upper ($n_\emr{up}$) and lower ($n_\emr{low}$) levels of the studied
line
\eq{ \frac{n_\emr{up}}{n_\emr{low}} = \frac{g_\emr{up}}{g_\emr{low}}
\exp\left[ - \frac{h\nu_l}{kT_\emr{ex}} \right].
  \label{eq:tex} }
When the molecules are in thermal equilibrium with their environment, the
temperature $T_\emr{ex}$ is equal to the gas kinetic temperature.  The
kinetic temperature is not known and must be estimated.

In the previous equations, the physical characteristics of the gas ($N$,
$T_\emr{ex}$, $\sigma_V$, and $\Delta_{V}$) depend on the specific line of
sight on the sky. Moreover, while observers try to get a uniform noise when
observing the source, this is never perfect and it is important to assume
that the noise standard deviation $\sigma_b$ also depends on the specific
line of sight on the sky. These considerations imply that $\alpha_l$,
$\tau_l$, $\nu^\emr{cent}_l$, $\Psi_l$, $s$, and $x$ will also depend on
the sky position.



\def\ValSigmab{100} 

\newcommand{\FigCRBDvSv}{%
  \begin{figure*}
    \centering{\includegraphics{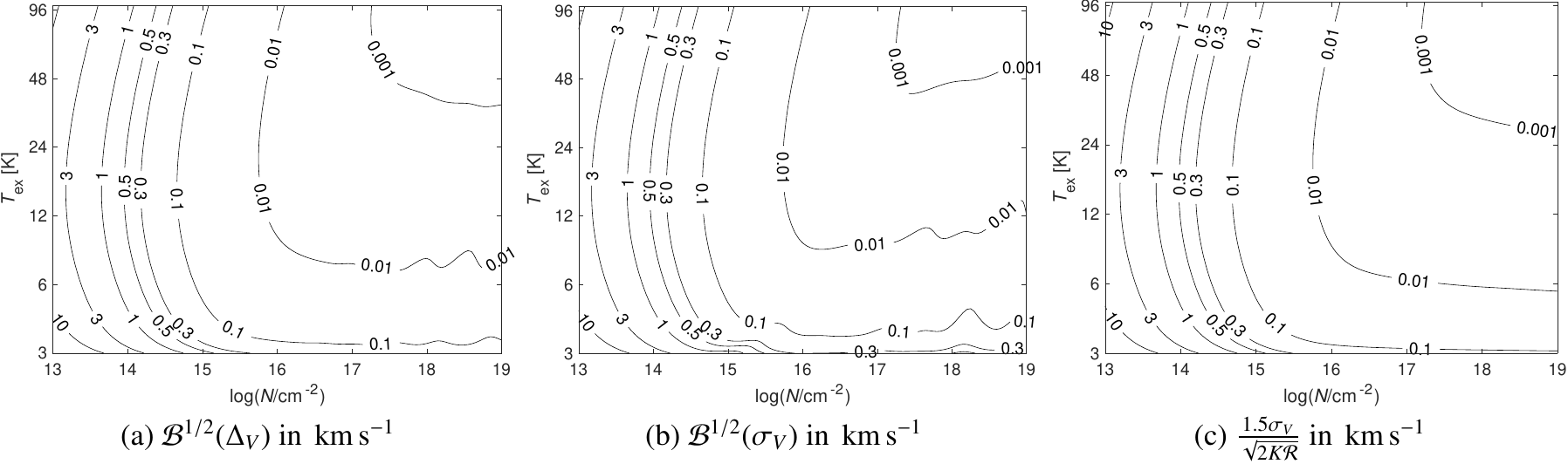}}
    \caption{Variations of the square root of the Cramer-Rao Bound
      (CRB) of the centroid velocity ($\Delta_V$, \textbf{left} panel) and
      velocity dispersion ($\sigma_V$, \textbf{middle} panel) in
      \unit{km\,s^{-1}} as a function of the column density and the
      excitation temperature. The \textbf{right} panel shows the variations
      of a function of the product of the number of channels $(K)$ and the
      signal-to-noise ratio $({\cal R})$.
      In all three cases, data are simulated assuming that \tcouz and \tcodu
      are measured and the unit of the image contours are \unit{km\,s^{-1}}.
      In this simulation, $\sigma_b=\ValSigmab\unit{mK}$,
      $\Delta_V=1.1\unit{km\,s^{-1}}$ and $\sigma_V=0.61\unit{km\,s^{-1}}$.}
    \label{Fig_CRB_sigmav_Deltav}
  \end{figure*}
}

\newcommand{\FigCRBTex}{%
  \begin{figure*}
    \centering{\includegraphics{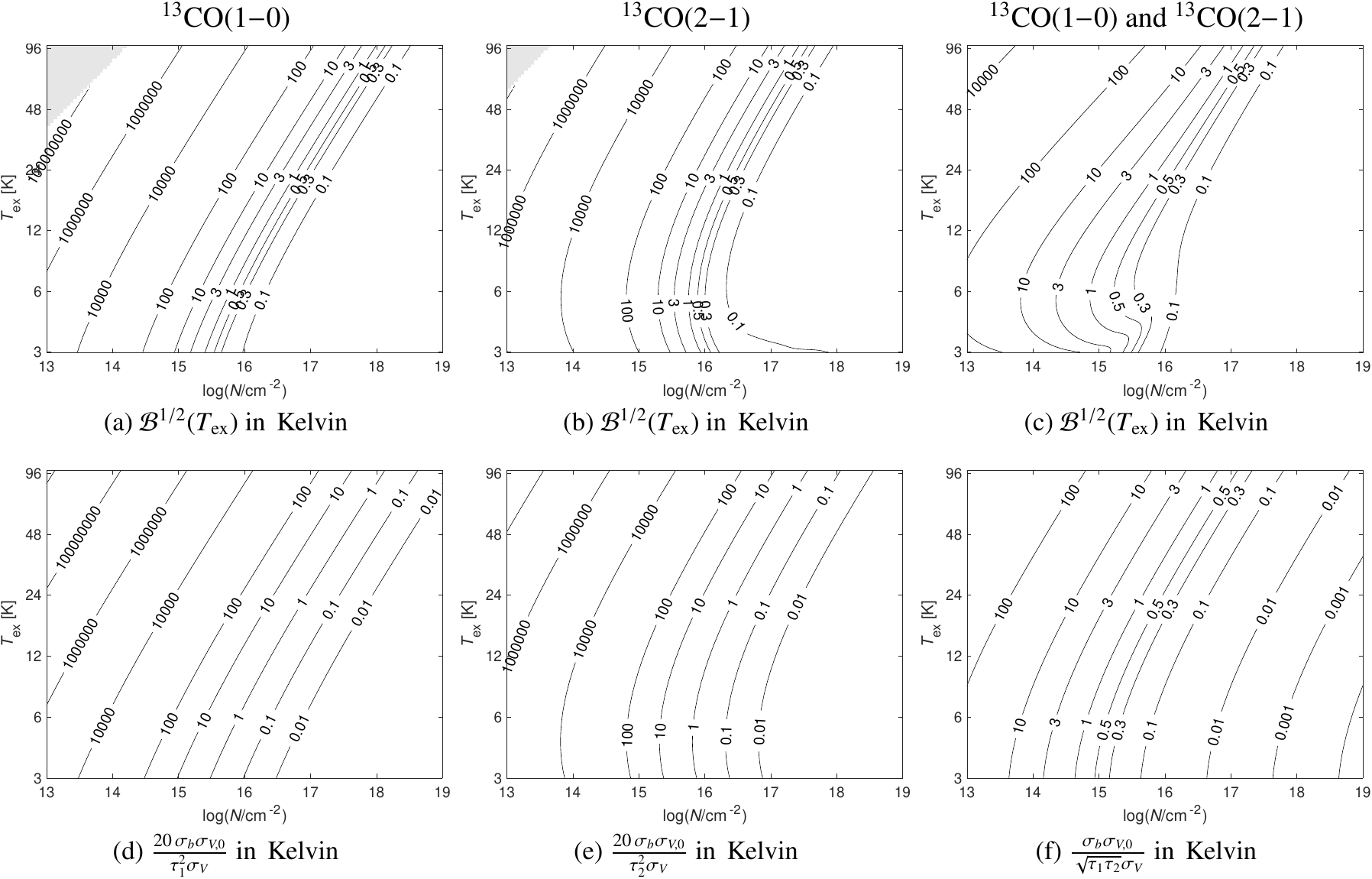}}
    \caption{\textbf{Top:} Variations of the square root of the CRB of
      $T_\emr{ex}$ in Kelvin as a function of the column density and the
      excitation temperature. \textbf{Bottom:} Functions of the line
      opacities.
      \textbf{Left:} Only \tcouz is analyzed. \textbf{Middle:} Only \tcodu is
      analyzed. \textbf{Right:} Both \tcouz and \tcodu are analyzed
      simultaneously.
      In (a) and (b) pixels in grey correspond to $T_\emr{ex}$ and $N$ values
      which lead to singular Fisher matrices.
      For this analysis, $\sigma_b=\ValSigmab\unit{mK}$,
      $\Delta_V=1.1\unit{km\,s^{-1}}$ and $\sigma_V=0.61\unit{km\,s^{-1}}$.
      The constant $\sigma_{V,0}=1\unit{km\,s^{-1}}$ is introduced to have
      expressions in (d-f) that depend on $\sigma_V$, but remain homogeneous
      to a temperature.}
    \label{Fig_CRB_T_m}
  \end{figure*}
}

\newcommand{\FigCRBlogN}{%
  \begin{figure*}
    \centering{\includegraphics{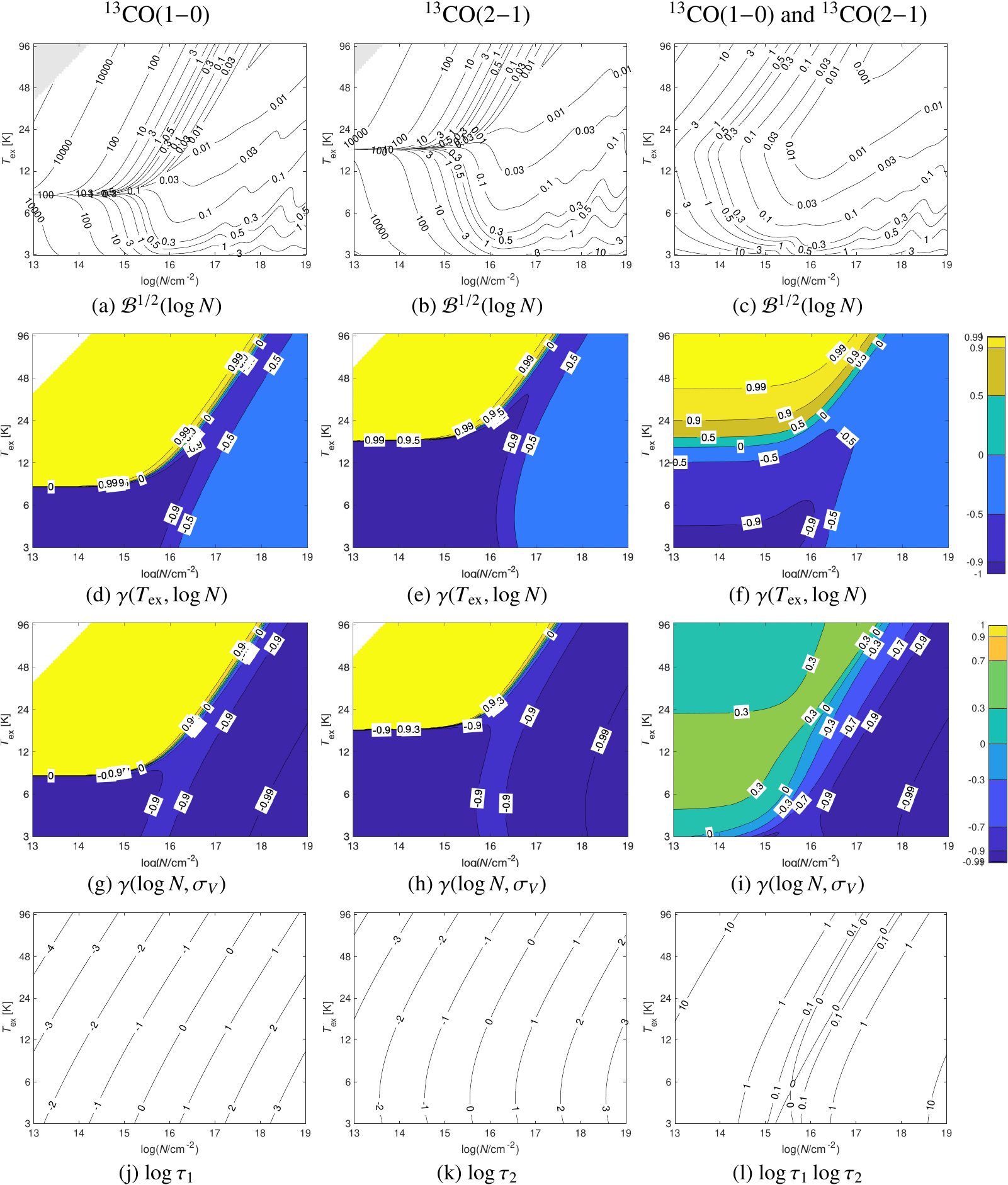}}
    \caption{\textbf{Top row:} Variations of the square root of the
      CRB of $\log N$ as a function of the column density and the excitation
      temperature. \textbf{Second and third row:} Correlation coefficients
      between efficient estimators of $(T_\emr{ex},\log N)$, and
      $(\log N, \sigma_V)$ in the second and third rows. respectively
      (defined in Eqs.~\eqref{eq_gamma_TN}
      and~\eqref{eq_gamma_Nsigmav}). \textbf{Bottom row:} Variations of
      functions of the opacities.
      \textbf{Left:} Only \tcouz is analyzed. \textbf{Middle:} Only \tcodu is
      analyzed. \textbf{Right:} Both \tcouz and \tcodu are analyzed
      simultaneously.
      In (a) and (b) pixels in grey correspond to $T_\emr{ex}$ and $N$ values
      which lead to singular Fisher matrices.
      For this analysis, $\sigma_b=\ValSigmab\unit{mK}$,
      $\Delta_V=1.1\unit{km\,s^{-1}}$ and $\sigma_V=0.61\unit{km\,s^{-1}}$.}
    \label{Fig_CRB_N_m}
  \end{figure*}
}

\newcommand{\FigCorrelations}{%
  \begin{figure}
    \centering{\includegraphics{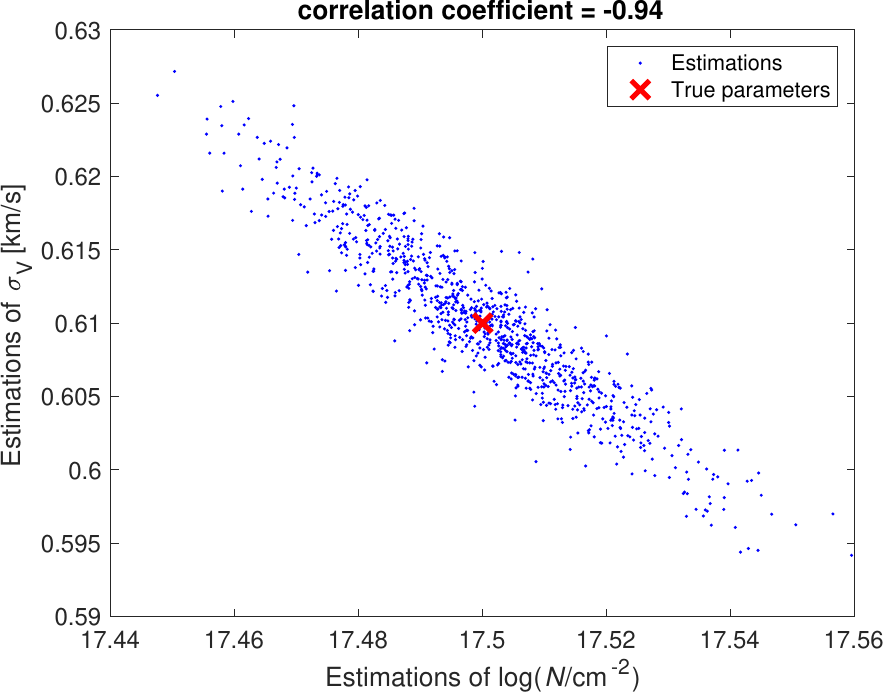}}
    \caption{Illustration of the correlation between $N$ and $\sigma_V$
      estimations when a single line (\tcouz) is available.  The blue
      points in the scatter plots show the estimations of $N$ and
      $\sigma_V$ obtained with a Monte Carlo simulation of individual
      spectra that share the same physical parameters and different
      realizations of a white Gaussian noise with standard deviation
      $\sigma_b=100 \unit{mK}$. The parameters are $T_\emr{ex}=18\unit{K}$,
      $N=10^{17.5}\unit{cm^{-2}}$, $\Delta_V=1.1\unit{km\,s^{-1}}$, and
      $\sigma_V=0.61\unit{km\,s^{-1}}$.}
    \label{fig_correlation}
  \end{figure}}

\newcommand{\FigMinNoiseOne}{%
  \begin{figure*}
    \centering{\includegraphics{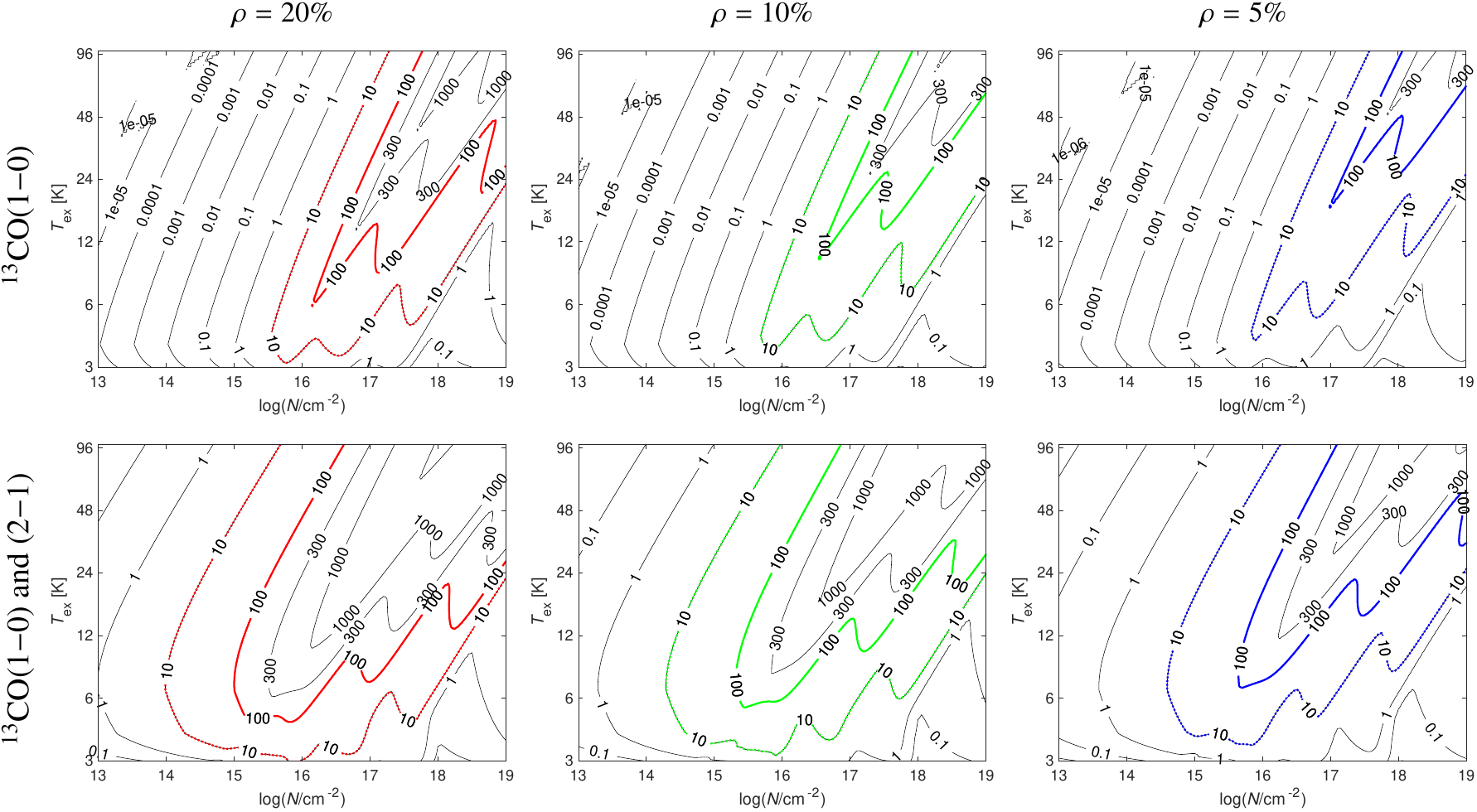}}
    \caption{Noise standard deviation $\sigma_{b,\rho}$ in $\unit{mK}$
      which ensures that relative precisions are better than $\rho\%$ (for
      details see Eq.~\eqref{eq_rho}).
      \textbf{Top:} A single line is analyzed. \textbf{Bottom:} Two lines
      \tcouz and \tcodu are analyzed.
      The contours for 10 and 100\unit{mK} are highlighted because these
      $\sigma_b$ values bracket the values reached during typical
      observations at the IRAM-30m.
      For this analysis, $\Delta_V$ is fixed at 1.1\unit{km\,s^{-1}}, but the
      computations are done for four different values of $\sigma_V$ (0.3,
      0.6, 1.3, and 2.0\unit{km\,s^{-1}}) and then projected on the
      $(N,T_\emr{ex})$ plane (see text for details).} %
    \label{Fig_quel_sigma_b}
  \end{figure*}
}

\newcommand{\FigMinNoiseTwo}{%
  \begin{figure*}
    \centering{\includegraphics{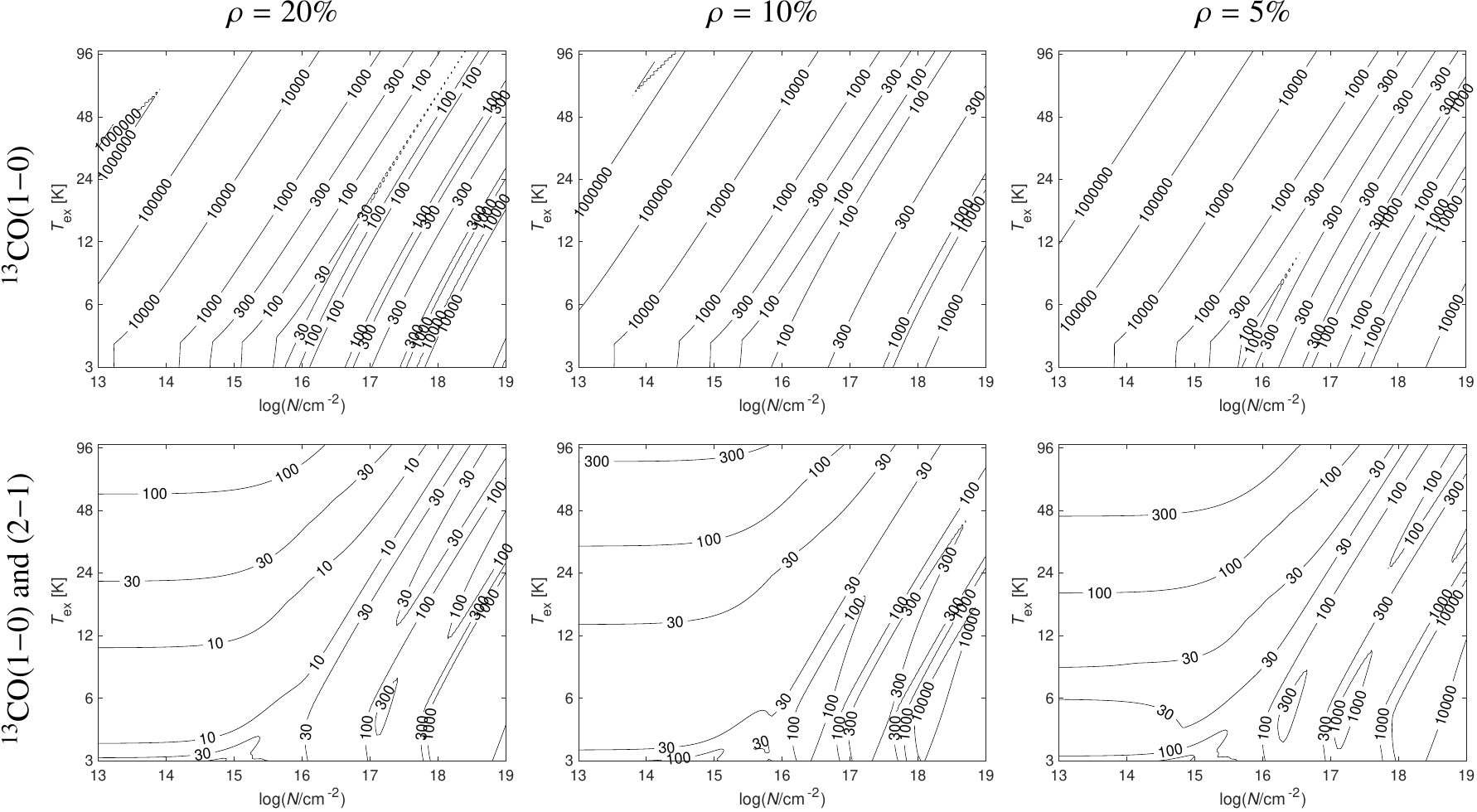}}
    \caption{Same as Figure~\ref{Fig_quel_sigma_b}, except that the
      contours show the variations of the peak-signal-to-noise ratio,
      ${\cal P}_\rho$, required to reach a given relative accuracy
      $(\rho\%)$.}
    \label{Fig_quel_sigma_b_cvt_en_P}
  \end{figure*}
}

\newcommand{\FigTwCOone}{%
  \begin{figure*}
    \centering{\includegraphics{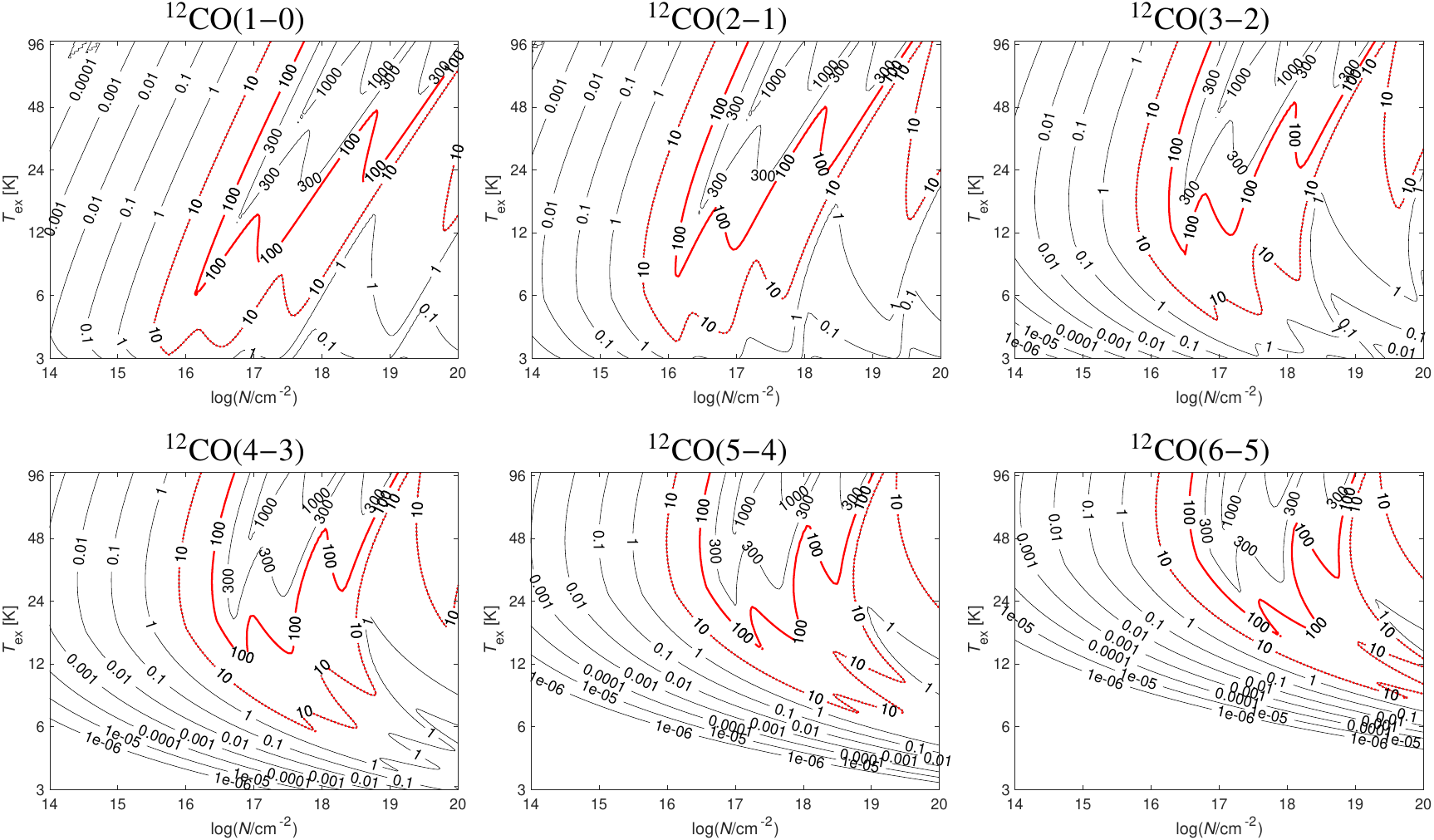}}
    \caption{Noise standard deviation $\sigma_{b,\rho}$ in $\unit{mK}$
      which ensures relative precisions better than 20\% when a single line
      of $^{12}$CO is analyzed. Other details are identical to
      Fig.~\ref{Fig_quel_sigma_b}.}
    \label{Fig_CRB_12CO_quel_line}
  \end{figure*}
}

\newcommand{\FigTwCOtwo}{%
  \begin{figure*}
    \centering{\includegraphics{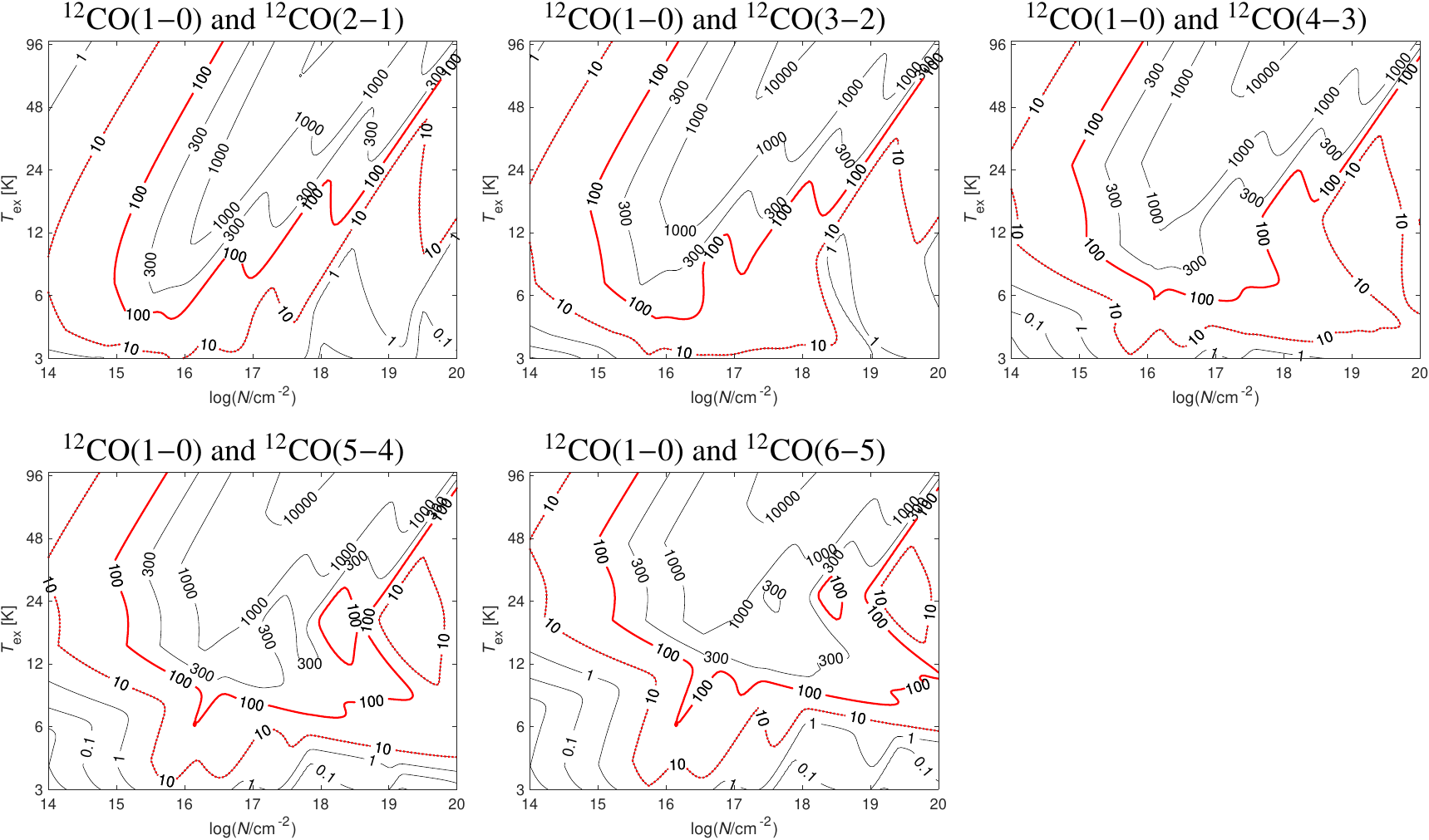}}
    \caption{Noise standard deviation $\sigma_{b,\rho}$ in $\unit{mK}$
      which ensures relative precisions better than 20\% when a couple of
      $^{12}$CO lines are observed: $J=1-0$ and a higher $J$ line. Other
      details are identical to Fig.~\ref{Fig_quel_sigma_b}.}
    \label{Fig_CRB_12CO_quel_line_avec_12co10}
  \end{figure*}
}


\section{Cramer-Rao Bound analysis}
\label{sec_CRB}

\FigCRBDvSv{}%

In this article, we aim at estimating the physical parameters of the LTE
model presented in section~\ref{sec_model} on the ORION-B data (see
section~\ref{sec_data}). Even when the LTE model is perfectly verified, the
presence of an additive Gaussian noise induces some uncertainty on the
estimation. For each physical parameter $\theta$ estimated as
$\widehat\theta$, the estimation error $(\widehat\theta-\theta)$ can be
quantified with the mean square error (MSE)
$\langle (\widehat\theta-\theta)^2\rangle$, where $\langle . \rangle$
represents the statistical mean over the different realizations of the
noise $b$. MSE can be estimated with Monte Carlo simulations, but the
result then depends on the choice of the implemented estimator. For
example, when MSE is large, one does not know whether it is due to the
choice of the estimator or to a lack of information in the data.
In estimation theory, the Fisher matrix allows to quantify the amount of
information in the considered problem. It provides a reference precision,
named Cramer-Rao Bound (CRB), which does not depend on the choice of a
specific estimator of the searched quantity, but only on the physical model
and the statistical properties of the noise \citep[see,
e.g.,][]{bonaca:18,espinosa:18}.

Mathematically speaking, the CRB noted ${\cal B}(\theta)$ is simply a lower
bound on the MSE of unbiased estimators (see Eq.~\eqref{eq_BCR}). Indeed,
the MSE is equal to the estimation variance
$\langle(\widehat\theta-\langle\widehat\theta\rangle)^2\rangle$ for an
unbiased estimator because the estimation MSE is in general equal to the
sum of its variance and its bias $(\langle\widehat\theta\rangle - \theta)$
squared, i.e.,
\begin{equation}
  \langle (\widehat\theta-\theta)^2\rangle
  = \langle(\widehat\theta-\langle\widehat\theta\rangle)^2\rangle
  + (\langle\widehat\theta\rangle - \theta)^2
\end{equation}
Therefore, a high CRB value implies that any unbiased estimator
$\widehat\theta$ will necessarily have a high dispersion around the true
value $\theta$. A high CRB can be understood as a lack of information in
the underlying model with respect to the considered level of noise. When it
occurs, one solution can be to introduce additional a priori knowledge or
to make another measurement with a better signal-to-noise ratio.
In contrast, a low CRB value does not necessarily imply that there exists
an unbiased estimator $\widehat\theta$ with a low dispersion around the
true value $\theta$. CRB is only a lower bound and it can be overly
optimistic. It is thus necessary to build an estimator, which can be tested
by comparing its variance with the CRB. If the estimator is unbiased and
its variance is equal to the CRB, then one knows that there does not exist
any better unbiased estimator.
In this section, we analyze the CRB (i.e. a bound on the variance of all
unbiased estimators), and in section \ref{sec_ML} we check with Monte Carlo
simulations that an efficient estimator (i.e., whose variance reaches the
CRB) exists.

We here compute the Fisher matrix and the associated CRB precisions for the
LTE radiative transfer. We then study the variations of these reference
precisions for the different unknown physical parameters ($\Delta_V$,
$\sigma_V$, $T_\emr{ex}$, and $N$) as a function of the excitation
temperature and column density. We finally use the CRB precision to answer
two questions. First, what is the maximum noise tolerable to get a given
relative precision on these parameters? We here compare the cases where
only one (\tcouz) or two (\tcouz and \tcodu) lines are available. Second,
which $^{12}$CO line should be observed to improve the precision achieved
when only \dcouz observations are available?

\subsection{Computing the CRB from the Fisher matrix for a single line and
  a single velocity component}
\label{sec_Fisher_matrix}

For a given line $l$, a sampled version of Eq.~\eqref{eq_x} can be written
over $K$ discrete frequency channels as
\eq{ \forall n\in\{1,...,K\} \quad x_{n,l}=s_{n,l}+b_{n,l}.
  \label{eq_x_n} }
When $b$ is a centered white Gaussian noise of standard deviation
$\sigma_{b,l}$, and the physical model $s$ is expressed as a function of a
set of unknown parameters $(\theta_i)$, the Fisher matrix $\bI_F$, which
represents the amount of information provided by line $l$, can simply be
computed as~\citep{sto05}
\eq{ \forall (i,j) \quad \left[\bI_F\right]_{ij}=\frac{1}{\sigma_{b,l}^2}
  \sum_{n=1}^K \frac{\partial s_{n,l}}{\partial \theta_i}\frac{\partial
    s_{n,l}}{\partial \theta_j},
  \label{eq_If} }
where $[\bA]_{ij}$ stands for the term $ (i,j) $ of the matrix $\bA$.

In our case, the physical model for $s$ will be the LTE radiative transfer
introduced in Sect.~\ref{sec_model} and the vector of unknown parameters is
$\btheta=[T_\emr{ex},\, \log N,\,\Delta_{V},\,\sigma_{V}]^T$, which we also
write $\btheta=[\theta_1,\, \theta_2,\, \theta_3,\, \theta_4]^T$ to
simplify the expression of the Fisher matrix in Eq.~\ref{eq_If}. In this
vector of parameters, we chose to analyze the precision of the
logarithm\footnote{In this paper, the notation $\log$ refers to the
  logarithm in base 10.} of the column density $N$ instead of directly
analyzing the precision of $N$, because the column density can vary over
orders of magnitudes in Giant Molecular Clouds.

It can be shown \citep{gar95} that the variance of any unbiased estimator
$\widehat \theta_i$ (here $\widehat T_\emr{ex}$, $\log \widehat N$,
$\widehat \Delta_V$ or $\widehat \sigma_V$) is bounded by
\eq{ \var (\widehat \theta_i) \geq {\cal B}(\theta_i)=[\bI_F^{-1}]_{ii}.
  \label{eq_BCR} }
Each diagonal term of the inverse of the Fisher matrix
${\cal B}(\theta_i)=[\bI_F^{-1}]_{ii}$ is called the Cramer Rao bound of
the corresponding parameter. We will note them $\CRB(T_\emr{ex})$,
$\CRB(\log N)$, $\CRB(\Delta_V)$ and $\CRB(\sigma_V)$.
These CRBs do not depend on the choice of the estimation algorithm
$\widehat \btheta$ and are usually asymptotically reached by the maximum
likelihood estimator \citep{gar95}. \textit{The CRB can thus be considered
  as a reference precision of the estimation problem.}

The calculation of the gradients
$\left(\frac{\partial s_{n,l}}{\partial \theta_i}\right)_{i=1,2,3,4}$ is
detailed in appendix~\ref{sec_Fisher}.

\subsection{Generalization to two lines and two velocity components}
\label{sec_two_components}

\FigCRBTex{}%
\FigCRBlogN{}%

We will use the CRB analysis on the case where we observe two different
lines $ (l\in\{1,\,2\}) $ of the same species. We will assume that these
lines are well separated in frequency so that their frequency supports are
disjoint
\eq{ x_{n,l}=s_{n,l}+b_{n,l} \quad \forall n\in\{1,...,K\} \quad \forall
  l\in\{1,\,2\}.
  \label{Eq_line_l} }
The Fisher matrix of the set $\left(x_{n,l}\right)$ is simply the sum of
the Fisher matrices of each transition because we assume that the unknown
parameters $\btheta$ are identical for the two lines.

We will also use the CRB framework in the case where each observed line is
emitted from two independent velocity components, i.e., from two gas
components characterized by different values of the unknown parameters
($\btheta_m$ with $m\in\{1,2\})$. Equation~\ref{Eq_line_l} that encodes the
spectrum for line $l$ can then be written as
\eq{ x_{n,l}=S_{n,l}+b_{n,l} \quad \forall n\in\{1,...,K\} \quad \forall
  l\in\{1,\,2\}, }
\eq{ \mbox{where} \quad S_{n,l}=s_{n,l}(\btheta_1)+s_{n,l}(\btheta_2).  }
This means that the composite line profile is considered as the simple sum
of two velocity components that do not radiatively interact. This
assumption is only correct if the two velocity components are sufficiently
separated in velocity. This case with two components is the most complex
model we will study in this paper. In this case, the number of unknown
parameters is 8 (instead of 4) and thus the size of the Fisher Matrix is
$8\times 8$ (instead of $4\times 4$).

\subsection{CRB variations as a function of $T_\emr{ex}$ and $N$}
\label{sec_CRB_varitions}

As the inversion of the Fisher matrix is done numerically, we do not have a
simple explicit expression of the CRBs. In this section, we thus
empirically analyze their evolution as a function of the physical
properties of the analyzed medium in a particular case taken from the
ORION-B project.  To generate figures~\ref{Fig_CRB_sigmav_Deltav},
\ref{Fig_CRB_T_m}, and \ref{Fig_CRB_N_m}, we assume that the two measured
lines are \tcouz and \tcodu. The two corresponding opacities are noted
$\tau_1$ and $\tau_2$. The number of samples is $K=80$ for each line at a
spectral resolution of $0.5\unit{km\,s^{-1}}$. Only one velocity component
is assumed in the remainder of this section. The amount of noise is fixed
and identical for both lines at $\sigma_{b,1}=\sigma_{b,2}=100\unit{mK}$.

The values of the Cramer-Rao Bounds of the unknown parameters (i.e.,
$T_\emr{ex}$, $\log N $, $\Delta_V$ and $\sigma_V$) are then computed as a
function of the values of $T_\emr{ex}$ and $N$. The Fisher matrices are
computed following Eq.~\eqref{eq_If}, and then numerically inverted to
obtain ${\cal B}(\theta_i)=[\bI_F^{-1}]_{ii}$. The excitation temperature
$T_\emr{ex}$ is sampled logarithmically between $3\unit{K}$ and
$99\unit{K}$, the column density $N$ is sampled logarithmically between
$10^{13}\unit{cm^{-2}}$ and $10^{19}\unit{cm^{-2}}$, and the other two
parameters are kept constant at $\Delta_V=1.1\unit{km\,s^{-1}}$ and
$\sigma_V=0.61\unit{km\,s^{-1}}$ (arbitrarily chosen). This leads to
figures~\ref{Fig_CRB_sigmav_Deltav}, \ref{Fig_CRB_T_m}, and
\ref{Fig_CRB_N_m}, where $N$ and $T_\emr{ex}$ varies horizontally, and
vertically, respectively.  In these figures, the variations of
${\cal B}^{1/2}(\theta_i)$ are shown instead of the variations of
${\cal B}(\theta_i)$ because the square root of the CRB is homogeneous to
the estimation standard deviation. It can thus be interpreted as errorbars
on the estimated parameter $\theta_i$. Varying $T_\emr{ex}$ and $N$ changes
not only the signal-to-noise ratio, but also the amount of information
measured by the Fisher matrix because of the non linearity in the
radiative transfer equation.

\subsection{Precision of the estimation of the centroid velocity $\Delta_V$
  and the associated velocity dispersion $\sigma_V$}

Figure~\ref{Fig_CRB_sigmav_Deltav}(a-b) shows variations of
$\CRB^{1/2}(\Delta_V)$ and $\CRB^{1/2}(\sigma_V)$. For
$N\geq 10^{16}\unit{cm^{-2}}$ and $T_\emr{ex}\geq 12\unit{K}$, the square
root of both CRBs are smaller than $0.01 \unit{km\,s^{-1}}$. This means
that any efficient unbiased estimator will have a small dispersion around
the actual values. This can be written
$\widehat\Delta_V=1.10 \pm 0.01\unit{km\,s^{-1}}$ and
$\widehat\sigma_V=0.61 \pm 0.01\unit{km\,s^{-1}}$.

Figure~\ref{Fig_CRB_sigmav_Deltav}(c) shows a function of $K{\cal R}$,
where ${\cal R}$ is the signal-to-noise ratio defined by
\eq{{\cal R}=\frac{\sum_{n=1}^K (s_{n,1}^2+s_{n,2}^2)}
  {K(\sigma_{b,1}^2+\sigma_{b,2}^2)}.}
This expression is used in signal processing to quantify the
signal-to-noise ratio on the ``energy'' of the signal. In our case, we
empirically find that, as a rule of thumb
\eq{ \CRB^{1/2}(\Delta_V) \simeq \CRB^{1/2}(\sigma_V) \simeq \frac{1.5
    \sigma_V}{\sqrt{2K{\cal R}}}.
  \label{eq_approx_CRB_Deltav_sigmav} }
While the dependency on $\sigma_V$ is not presented in
Fig.~\ref{Fig_CRB_sigmav_Deltav}, we checked that
Eq.~\eqref{eq_approx_CRB_Deltav_sigmav} remains valid when $\sigma_V=0.3$,
$1.31$, and $2\unit{km\,s^{-1}}$. The estimation precision on $\Delta_V$
and $\sigma_V$ depends on the signal-to-noise ratio and on $\sigma_V$. This
is expected because of the similarity with the problem of delay estimation
in radar for which \citet{che89} obtained an analytic formulation similar
to Eq.~\eqref{eq_approx_CRB_Deltav_sigmav}.

\subsection{Precision of the estimation of the excitation temperature
  $T_\emr{ex}$}

In this section, we start to quantitatively evaluate the gain in precision
when two lines are observed instead of a single
one. Figure~\ref{Fig_CRB_T_m} compares the variations of
$\CRB^{1/2}(T_\emr{ex})$ when only \tcouz or \tcodu is available to
constrain the excitation temperature, and when both lines are available. To
interpret this figure, we first mention that, for low column densities, the
uncertainty quickly increases leading to large values of the CRB,
especially for $N < 10^{16}$ cm$^{-2}$. While the variation of the CRB as a
function of the excitation temperature for a given column density is
monotonous in the considered range for the \tcouz{} line, it shows a
different behaviour for the \tcodu{} line, with a minimum near 6\,K, and an
increase of the CRB for lower values of the excitation temperature.  This
different behaviour is related to higher energy of the upper state of the
$2-1$ transition. The emerging \tcodu{} signal, which is proportional to
the population of the upper level of the transition, approaches zero and
becomes close to the noise level.

For the considered example, the analysis of a single line (see
Fig.~\ref{Fig_CRB_T_m}~(a-b)) will give a reference precision on
$T_\emr{ex}$ of $0.1\unit{K} \leq \CRB^{1/2}(T_\emr{ex})< 10\unit{K}$ for
typically $N>[10^{16}\,-\, 10^{17.5}]\unit{cm^{-2}}$. The dependence on
$T_\emr{ex}$ is such that the same CRB is also reached at higher column
densities for higher values of $T_\emr{ex}$. This behavior of the CRB can
be qualitatively understood as resulting from the increase of the line
opacity. When the opacity becomes larger than about 3, the peak temperature
only depends on $T_\emr{ex}$ as the factor $[1-\exp(-\Psi(\nu_l^{cent}))]$
in Eq (4) approaches unity.  The CRB almost linearly depends on
$\log T_\emr{ex}$ above $6\unit{K}$. The analysis of two lines (see
Figure~\ref{Fig_CRB_T_m}c) greatly improves the situation. One reaches the
same precision on $T_\emr{ex}$ for column densities that are between one
and two orders of magnitude lower, i.e.,
$0.1\unit{K} \leq\CRB^{1/2}(T_\emr{ex})<10\unit{K}$ for typically
$N>[10^{14}\,-\, 10^{16.5}]\unit{cm^{-2}}$. Here again the precision almost
linearly depends on $\log T_\emr{ex}$ above $6\unit{K}$.

The second row of Figure~\ref{Fig_CRB_T_m} shows functions of the
opacities. Trying for several values of $\sigma_V$, we empirically obtain
\eq{ \CRB^{1/2}(T_\emr{ex}) \simeq \frac{20\,\sigma_b \,\sigma_{V,0}}
  {\tau_l^2 \,\sigma_V},
  \label{eq_CRB_T_m_1} }
when a single line is available, either \tcouz or \tcodu. In this equation,
$\sigma_{V,0}=1\unit{km\,s^{-1}}$ is a constant fixed so that the
expression depends on $\sigma_V$, but remains homogeneous to a temperature.
When these two lines are available, we obtain
\eq{ \CRB^{1/2}(T_\emr{ex}) \simeq \frac{\sigma_b \,
    \sigma_{V,0}}{\sqrt{\tau_1 \tau_2}\,\sigma_V}.
  \label{eq_CRB_T_m_2} }
Hence, according to Eq.~\eqref{eq_CRB_T_m_1} and Eq.~\eqref{eq_CRB_T_m_2},
when opacities are close to one, the gain in precision (in standard
deviation) is around 20 when one observes two lines of the same species
instead of a single one. These relations suggest that the parameters that
control the difficulty of the estimation problem are the amount of noise
$\sigma_b$, the velocity dispersions $\sigma_V$, and the opacities.

\FigCorrelations{}%
\FigMinNoiseOne{} %
\FigMinNoiseTwo{} %

\subsection{Precision of the estimation of the column density $N$}
\label{sec_CRB_N}

The top row of Fig.~\ref{Fig_CRB_N_m} shows that the precision on the
estimation of $N$ has a complex behavior when only one line is
available. To interpret this, we note that even efficient estimators of
$\btheta$ have correlated components described by the correlation
coefficients of the Fisher matrix.  The correlation coefficient between
$T_\emr{ex}$ estimations and $\log N$ estimations is given by
\eq{ \gamma(T_\emr{ex}, \log N)= \frac{\CRB(T_\emr{ex}, \log
    N)}{\CRB^{1/2}(T_\emr{ex}) \CRB^{1/2}(\log N)},
  \label{eq_gamma_TN} }
where $\CRB(T_\emr{ex}, \log N)=[\bI_F^{-1}]_{12}$ is the non-diagonal
element of the inverse Fisher matrix, see Eq.~\eqref{eq_If}. We also
introduce the correlation coefficient between $\log N$ estimations and
$\sigma_V$ estimations
\eq{ \gamma(\log N,\sigma_V)= \frac{\CRB(\log N,\sigma_V)}{ \CRB^{1/2}(\log
    N) \CRB^{1/2}(\sigma_V)}
  \label{eq_gamma_Nsigmav} } where
$\CRB(\log N,\sigma_V)=[\bI_F^{-1}]_{24}$, see Eq.~\eqref{eq_If}.

Correlation coefficients are built such that their value ranges from -1 to
1. As long as $|\gamma|<1$, CRBs remain finite and thus estimating
parameters usually remains possible.  There is a complete ambiguity between
estimations of the pairs ($T_\emr{ex}$ and $\log N$) or ($\log N$ and
$\sigma_V$), only when values of $|\gamma| =1 $. In this case, the variance
of these estimations becomes infinite.
Figure~\ref{fig_correlation} shows a simulated example where $\log(N)$ and
$\sigma_V$ can be accurately estimated even though they are highly (but not
completely) anti-correlated. Starting from a modeled spectrum with
$\log(N/\unit{cm^{-2}}) = 17.5$, $\sigma_V = 0.61$\unit{km\,s^{-1}}, and
$T_\emr{ex} = 18\unit{K}$, we built one thousand realizations of the
observed spectrum with a Monte Carlo simulation, and we fitted the LTE
model using the estimator proposed in Sect.~\ref{sec_ML}. This Monte Carlo
simulation allows us to numerically estimate the standard deviation on the
estimated parameters and the correlation coefficient between $\log(N)$ and
$\sigma_V$. This coefficient is -0.94, implying that the parameters are
highly anti-correlated. However, the standard deviation on the
$\log(N/\unit{cm^{-2}}$ and $\sigma_V$ estimations are 0.017 and
$0.005\unit{km\,s^{-1}}$, respectively. This corresponds to typical
relative errors of 4.0 and 0.8\%, respectively. Hence some high
(anti-)correlation does not necessarily imply that the model parameters can
not be estimated, in contrast with a widespread intuition. While we
illustrated this property with a given estimator, this statement is true
for the CRB analysis. This emphasizes another of its interests. It provides
standard deviations and coefficient of correlations without requiring to
implement any Monte Carlo simulation.

The second and third row of Fig.~\ref{Fig_CRB_N_m} show important
ambiguities (i.e., correlation coefficients close to one or minus one)
between estimations of $\log N$ and $T_\emr{ex}$ and even more ambiguities
between estimations of $\log N$ and $\sigma_V$ (in particular for large
values of $N$).
When a single line is observed (Fig.~\ref{Fig_CRB_N_m}~d\,e,\,g,\,h),
$|\gamma(T_\emr{ex}, \log N)|$ and $|\gamma(\log N,\sigma_V)|$ are mostly
larger than 0.9 for small $N$ (in Fig.~\ref{Fig_CRB_N_m}~d-e yellow pixels
correspond to $\gamma>0.99$ and in g-h to $\gamma>0.9$). For high $N$, the
ambiguity with $T_\emr{ex}$ decreases, but not the one with $\sigma_V$ (in
Fig.~\ref{Fig_CRB_N_m}~d-e light blue values are $-0.5<\gamma<0$ while in
Fig.~\ref{Fig_CRB_N_m}~g-h dark blue values correspond to $\gamma<-0.9$
and $\gamma<-0.99$). The horizontal asymptote on the left of the CRB maps
corresponds to a very sharp change of sign of correlation coefficients
$\gamma$.
Figure~\ref{Fig_CRB_N_m}~f and i show that, although some ambiguities
remain between $\log N$ and $\sigma_V$ for high $N$ and small $T_\emr{ex}$,
having two lines mitigates these ambiguities in most cases.

As a rule of thumb, with a single line (see Fig.~\ref{Fig_CRB_N_m}~a-b),
$\CRB^{1/2}(\log N)<0.1$ for $N>[10^{16}\,-\,10^{17}]\unit{cm^{-2}}$
(depending on $T_\emr{ex}$), and $T_\emr{ex}\geq [6-12]\unit{K}$ (depending
on $N$).
With two lines (Fig.~\ref{Fig_CRB_N_m}~c), the situation greatly improves:
$\CRB^{1/2}(\log N)<0.1$ for $N>[10^{15}\,-\,10^{17}]\unit{cm^{-2}}$
(depending on $T_\emr{ex}$) and $T_\emr{ex}\geq 6\unit{K}$.
Figures~\ref{Fig_CRB_N_m}~a,\,b,\,c also show local minima of the
$\CRB^{1/2}(\log N)$ when $T_\emr{ex}$ increases and
$N\geq 10^{17}\unit{cm^{-2}}$, and when $N$ increases and
$T_\emr{ex} \geq 6\unit{K}$. To interpret these, the last row of
Fig.~\ref{Fig_CRB_N_m} shows functions of the opacities. Comparing these
with the variations of $\CRB^{1/2}(\log N)$ shows that the smallest values
of $\CRB^{1/2}(\log N)$ (i.e., the best achievable precision) are mainly
located at the area where $\tau_1$ and $\tau_2$ are close to 1. With two
lines (see Fig.~\ref{Fig_CRB_N_m}l), the best precision is when $\tau_1<1$
and $\tau_2>1$.

\subsection{Maximum noise tolerable to get a given relative precision on
  the different parameters}
\label{sec_CRB_relative}

In the previous section, the standard deviation of the noise $\sigma_b$ was
fixed to $100\unit{mK}$. Conversely, we now derive the amount of noise that
guarantees a given relative CRB precision for $T_\emr{ex}$, $\log N$,
$\sigma_V$ and $\Delta_V$.  We compute $\sigma_{b,\rho}$ the maximal value
of $\sigma_b$ that satisfies the following inequalities
\eq{
  \begin{array}{cc}
    \CRB^{1/2}(T_\emr{ex})/T_\emr{ex} \leq\rho, & \CRB^{1/2}(\log N)\leq\rho, \\
    \CRB^{1/2}(\sigma_V)/\sigma_V \leq\rho,     & \CRB^{1/2}(\Delta_V)/\sigma_V \leq\rho,
  \end{array}
  \label{eq_rho} }
where $\rho$ is a fixed threshold.  In other words, instead of analyzing
the precision for a given amount of noise, one can also analyze the
tolerable level of noise to ensure an intended precision (herein described
by $\rho$). Such an analysis will be useful to design an observation
program and optimize the telescope time needed to reach the scientific
goal.

Up to this point of the paper, we checked the variations of the quantities
as a function of $T_\emr{ex}$ and $N$ with fixed values of $\Delta_V$ and
$\sigma_V$.  Figure~\ref{Fig_quel_sigma_b} shows the variations of
$\sigma_{b,\rho}$ as a function of $T_\emr{ex}$ and $N$. As the computation
of $\sigma_{b,\rho}$ includes the computation of maximum values, it is
possible to make the computations in three dimensions (with varying values
of $T_\emr{ex}$, $N$, and $\sigma_V$), and to project these on the
$(T_\emr{ex},N)$ plane. That is what is shown in
Fig.~\ref{Fig_quel_sigma_b} for different values of the relative precision
$\rho$ (5, 10, and 20\%).

The IRAM-30m time estimator for the EMIR
receivers\footnote{\url{http://www.iram.es/nte/}} indicates that we can
achieve a sensitivity of 100\,mK in $30$ to $120\unit{s}$ at 110 and
220\,GHz for a spectral resolution of 0.5\unit{km\,s^{-1}}. Similarly, we
can achieve a sensitivity of 10\,mK in 1 to 3 hours depending on the
frequency and the observing mode (frequency or position switching). The
colored contours thus correspond to the ``fast/slow'' acquisition mode at
the IRAM-30m. The comparison between the top and bottom lines allows us to
see the gain in precision when analyzing the two lowest $J$ lines of
$^{13}$CO instead of a single one.  In particular, the surface of reachable
combinations of column density and excitation temperature more than doubles
when analyzing two lines.

Instead of analyzing the level of noise, one can also analyze the
peak-signal-to-noise ratio defined for one transition $l$ by
\eq{ {\cal P}_l=\frac{\max_n s_{n,l}}{\sigma_{b,l}}
  \label{eq_P_l} } and for two transitions by
${\cal P}=\max_{l=1,2} {\cal P}_l$.
Figure~\ref{Fig_quel_sigma_b_cvt_en_P} shows the variations of minimum
peak-signal-to-noise ratio ${\cal P}_{\rho}$ for similar conditions as in
Fig.~\ref{Fig_quel_sigma_b}. Analyzing only the \tcouz line requires at
least a signal-to-noise ratio of 100 to get a relative precision of
20\%. Adding the \tcodu line in the analysis reduces the minimum
signal-to-noise ratio by a factor up to 10 to reach the same relative
precision. The required signal-to-noise ratio increases at high column
densities because the lines become optically thick, and at a combination of
low column density and high excitation.

\subsection{How to complement \dcouz observations?}
\label{sec_complet_dcouz}

All the previous analyses were done for the $^{13}$CO isotopologue because
it enabled us to study the case of low $J$ lines that experience the
transition from optically thin to thick regime over the range of column
densities and excitation temperatures that are found in molecular
clouds. However, the targeted transition when observing the molecular gas
of a new astronomical source is usually \dcouz because it is the strongest
line in the easily observable 3mm atmospheric window
\citep{wilson:70,pety:17}.

\FigTwCOone{}%
\FigTwCOtwo{}%

We here ask two questions. First, what is the best $J$ line to observe to
reach a relative precision of 20\% on all the estimated parameters?
Figure~\ref{Fig_CRB_12CO_quel_line} shows the variations of the maximum
noise $\sigma_{b,\rho}$ when a single line is observed among the first six
rotational transitions of $^{12}$CO. A global pattern is seen, especially
for the higher energy transitions \dcoqt, \dcocq, \dcosc, For instance, if
$N$ lies in the interval $[10^{17},\,10^{19}]\unit{cm^{-2}}$ and
$T_\emr{ex}> 24\unit{K}$, the \dcosc line seems the best choice (from a CRB
point of view) because it allows to reach $20\%$ accuracy over this broad
range of parameters for a noise level of 100\,mK.  However this line is not
easy to access from ground based telescopes because of the limited
atmospheric transmission at the line frequency of 690\,GHz.

Second, what is the best $J$ line to be observed to complement the $J=1-0$
line to reach the same relative precision of $20\%$?
Figure~\ref{Fig_CRB_12CO_quel_line_avec_12co10} seems to indicate that
observing \dcosc would be the most useful as it would allow to tolerate a
noise level of $\sigma_b=300\unit{mK}$ and keep a good precision for $N$ in
the interval $[10^{16.0},\, 10^{18.5}]\unit{cm^{-2}}$.
We stress that this result only applies to the case where all transitions
have the same excitation temperature. In practice, deviations from a
Boltzmann population may be present leading to different excitation
temperatures for the $^{12}$CO transitions \citep{vandertak:07} because of
the higher critical densities of the higher-$J$ transitions. Nevertheless,
the usefulness of mildly excited lines remains valid. Non-LTE approaches
will be developed in the future that will provide a quantitative assessment
of the diagnostic power of these lines.



\newcommand{\FigSpectra}{%
  \begin{figure*}
    \centering{\includegraphics{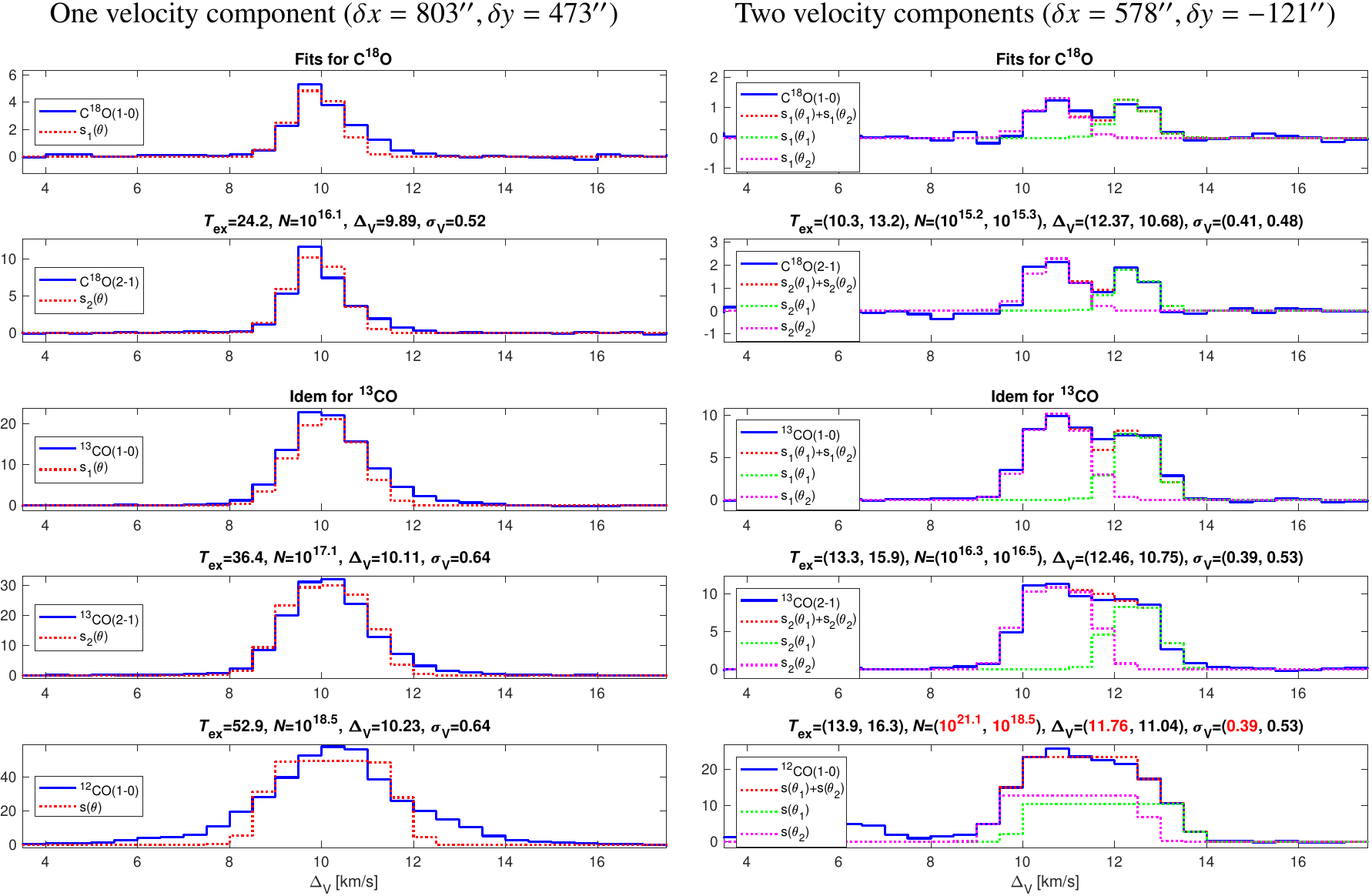}}
    \caption{Two examples of LTE fit of the CO isotopologues lines. In
      the titles, $T_\emr{ex}$ is expressed in Kelvin, $N$ in cm$^{-2}$,
      $\Delta_V$ in \emr{km\,s^{-1}}, and $\sigma_V$ in \emr{km\,s^{-1}}.
      The plain lines show the data, and the dotted ones show the fit
      results.  Values in red indicate estimations whose relative precision
      is larger than 20\%. The associated lines of sight can be localized in
      Fig.~\ref{Fig_Ima_in_pixels} (see red crosses).}
    \label{Fig_real_data_spectres}
  \end{figure*}
}

\newcommand{\FigResidualNoise}{%
  \begin{figure*}
    \centering{\includegraphics{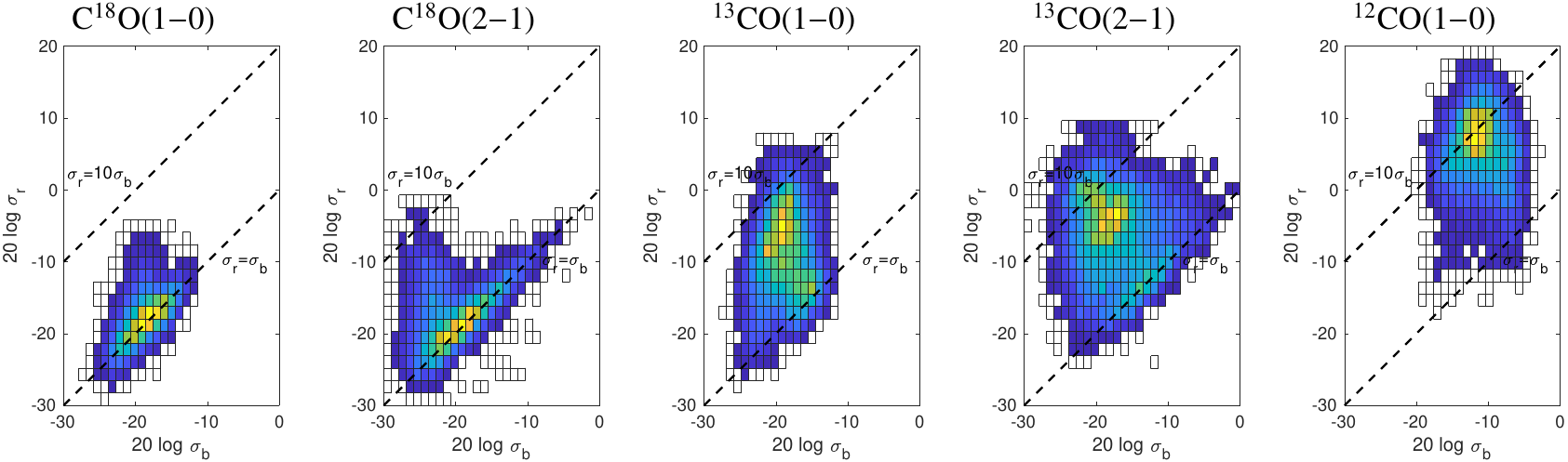}}
    \caption{Joint histograms of the standard deviations of the
      residuals $(\sigma_{r})$ and of the noise $(\sigma_{b})$ for all the
      studied lines. Standard deviations are expressed in Kelvin.  The dashed
      lines correspond to ratios 1 and 10.}
    \label{Fig_real_data_sigmab_Eb}
  \end{figure*}
}

\newcommand{\FigResidualEnergyOne}{%
  \begin{figure*}
    \centering{\includegraphics{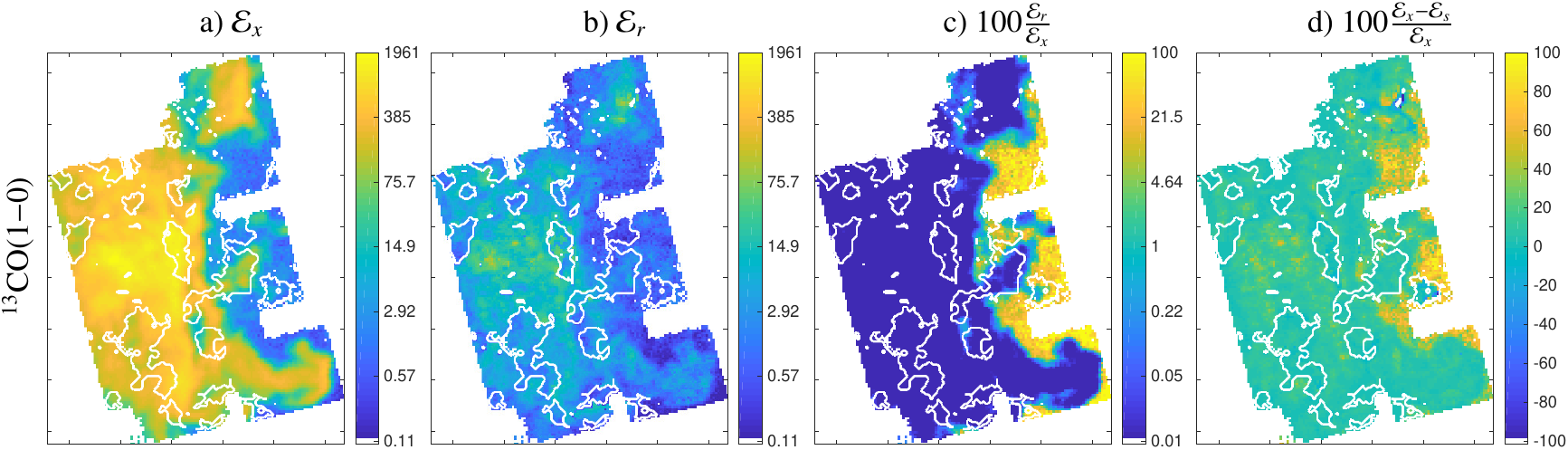}}
    \caption{Spatial variations of the observed spectrum ``energy''
      (\textbf{a}), of the residual ``energy'' (\textbf{b}), of their ratio
      in \% (\textbf{c}), and of the ratio of ``energy'' in \% that has not
      been modeled (\textbf{d}). The unit of the color look-up table is
      Kelvin$^{2}$.  White contours show the regions where two components
      have been detected.}
    \label{Fig_real_data_residus_tcouz}
  \end{figure*}
}


\section{Application to the ORION-B data}
\label{sec_real_data}

The CRB is only a lower bound on the variance of any unbiased
estimator. Once the order of magnitude of the CRBs have been analyzed, the
next step is to find a good estimator of physical parameters. In this
section, we first propose such an estimator and analyze its performance for
a realistic amount of noise (herein chosen to
$\sigma_b=\ValSigmab\unit{mK}$), before applying it to the ORION-B data.

\subsection{Proposed estimator}
\label{sec_ML}

The maximum likelihood estimator (MLE) is a good candidate because under
mild conditions, it reaches the CRB asymptotically, i.e., when
$\sigma_b\mapsto 0$ \citep{gar95}. Appendix~\ref{sec_MLE_definition}
details the computation of this estimator, its initialization, and the
iterative algorithm used to yield the estimation that is noted
$\widehat\btheta$. We also briefly discuss its computational efficiency.

Appendix~\ref{sec_MLE_performance} analyzes the performance of the proposed
estimator on simulated data. This appendix shows that this estimator
performs optimally for pairs of $(T_\emr{ex},N)$ values such that a
relative precision of reference is reached for all estimated parameters
(i.e., the conditions of Eqs.~\eqref{eq_rho} are satisfied with
$\rho=20\%$). Interestingly, this appendix also shows that the obtained
estimation $\widehat \btheta$ can be injected in the CRB computation to
detect whether or not this estimation is accurate.

\subsection{Estimation of the number of velocity components and
  initialization of the parameters}
\label{sec_init}

The number of velocity components is \textit{a priori} unknown. While we
could have tried to use our maximum likelihood estimator to fit the data
with either one or two components, we would then have had to devise a
statistical test to determine which assumption to choose.  Instead, it is
simpler to check for the presence of several local maxima in the spectrum
denoised with ROHSA, the technique mentioned in Sect.~\ref{sec_data}. In
particular, this allows us to have a spatially coherent detection of the
number of components.  The ROHSA algorithm is applied separately on the
\tcouz, \cdouz, and \dcouz lines.  If one of the denoised spectra for
\tcouz, \cdouz or \dcouz has at least two local maxima in the velocity
interval of interest $[8.25,14.25]\unit{km\,s^{-1}}$, we then fix the
number of velocity components to two for all three species.
The denoised spectra \tcouz, \cdouz and \dcouz are analyzed iteratively in
this order, and as soon as two components are selected based on one of the
three spectra, the velocities associated to the local maxima are used to
initialize the estimations of the velocity $\Delta_V$ of each component for
all three species. This ensures that the velocities of each component will
stay compatible for all three species during the fit. We analyze the
denoised spectra in the above order because the \tcouz{} line has both a
good signal-to-noise ratio, and moderate opacities. The \cdouz line
delivers a good information on the underlying velocity structure because it
is most often optically thin, but its limited signal-to-noise ratio may
hamper the $\Delta_V$ initialization.  The saturation that happens for the
\dcouz line also makes the determination of the $\Delta_V$ initializations
inaccurate. Finally, when all the three denoised spectra have only one
maximum, one component is independently fitted per species.

When initializing the parameters before maximizing the likelihood, we use
two different assumptions to help the algorithm to converge towards
reasonable solutions. First, when two components are detected, we use the
same $\Delta_V$ and $\sigma_V$ initializations for all the species so that
the estimated parameters for each component remain correctly paired among
the three species, as explained above. The white contours in
Fig.~\ref{Fig_real_data_results} delimit the regions where two local maxima
have been detected. Only $23\%$ of the field of view requires two velocity
components. As the estimations of $T_\emr{ex}$, $N$, and $\sigma_V$ are
highly correlated (see Sect.~\ref{sec_CRB_N}), we systematically search in
the 3D grid described in Sect.~\ref{sec_MLE_init} to initialize them.  We
stress that this is only during the initialization process that we use the
same values $\Delta_V$ and $\sigma_V$ for \tco and \cdo. The maximization
of the log-likelihood is done independently on each species, ensuring that
the estimations of $\Delta_V$ and $\sigma_V$ may take a different value for
each species.

Second, a single line of $^{12}$CO is available and it is quite optically
thick.  In our analysis of the CRB, we observed in
Figure~\ref{Fig_CRB_N_m}~(d-e and g-h) that the estimation of $\log N$ is
highly correlated with $\sigma_V$, when $N$ is high. This may imply some
degeneracy between the estimation of the velocity dispersion and the column
density.  To alleviate this issue, we first deal with \tcouz, \tcodu,
\cdouz, \cdodu, and we use the estimation of $\sigma_V$ obtained on
$^{13}$CO to fix $\sigma_V$ for the estimation of the other parameters
($T_\emr{ex}$, $\log N$ and $\Delta_V$) in the analysis of the \dcouz
line. If this assumption is false, the obtained estimations of the other
parameters will be biased. While this procedure is not ideal, we
empirically obtained estimations of $T_\emr{ex}$ and $\log N$ much closer
to physical intuition: in particular, the estimations of the column density
are 100 times too large when all four parameters are estimated. An analysis
of the impact of a potential incorrect value of $\sigma_V$ goes beyond the
scope of the present paper.

\FigSpectra{}%

\subsection{Detailed analysis of two lines of sight}

Figure~\ref{Fig_real_data_spectres} shows how the proposed estimator
succeeds to fit the C$^{18}$O, $^{13}$CO and $^{12}$CO low $J$ lines
towards two lines of sight in the studied field of view (see red crosses
Fig.~\ref{Fig_Ima_in_pixels}), at offsets $(803'',473'')$ and
$(578'',-121'')$. The spectra on the left column are modeled with a single
velocity component but they are asymmetric. This implies that our model is
not perfectly adequate because it assumes that the line profile is
symmetric. The issue is most problematic for $^{12}$CO, because the high
opacity and the complex underlying velocity field imply a more complex
profile with broad wings on each side of the line. A detailed solution for
this issue is beyond the scope of the present paper.

The spectra on the right column are well fitted with two different velocity
components. The estimations for the C$^{18}$O and $^{13}$CO species are
physically relevant because the two velocity components are well separated
in velocity. This is less obvious for $^{12}$CO, which presents a large
velocity overlap of the two components.

\subsection{Estimation of the quality of the fit and filtering out
  inaccurate estimations}

In this section, we first compute the ``energy'' and the standard deviation
of the fit residuals as two ways to quantify the quality of the fit. We
then explain how we will filter out inaccurate estimations from the
physical analysis.

After a fit, we can define three different ``energies'' in the sense of the
information theory.
\begin{itemize}
\item The ``energy'' of the measured signal is
  \eq{ %
    {\cal E}_{x_l}=\sum_n x_{n,l}^2.
    \label{eq_E_x}
  }
\item The ``energy'' of the estimated signal is
  \eq{ %
    \begin{array}{ccc}
      {\cal E}_{s_l}=\sum_n s^2_{n,l}(\widehat\btheta) %
      & \text{or} %
      & {\cal E}_{s_l}=\sum_n s^2_{n,l}(\widehat\btheta_1)+
        s^2_{n,l}(\widehat\btheta_2),
    \end{array}
    \label{eq_E_s}
  }
  depending of the number of estimated components.
\item The ``energy'' of the fit residual is
  \eq{ %
    {\cal E}_{r_l}=\sum_n r_{n,l}^2,
    \label{eq_E_b}
  }
  \eq{ %
    \begin{array}{ll}
      \text{where} & r_{n,l}=x_{n,l}-s_{n,l}(\widehat\btheta), \\
      \text{or}    & r_{n,l}=x_{n,l}-s_{n,l}(\widehat\btheta_1) - s_{n,l}(\widehat\btheta_2).
    \end{array}
    \label{eq_b}
  }
\end{itemize}
All the sums are computed on an interval of $10\unit{km\,s^{-1}}$ around the
maximum. The fit quality can then be quantified by comparing the ``energy''
in the residual with either the ``energy'' in the observed spectrum
$({\cal E}_{r_l}/{\cal E}_{x_l})$ or the difference of ``energy'' between
the observed and estimated signals
$(\{{\cal E}_{x_l}-{\cal E}_{s_l}\}/{\cal E}_{x_l})$. The former formula
tells us the fraction of the observed ``energy'' that has not been
fitted. The latter formula tells us whether the observed spectra has been
under-fitted (positive value) or over-fitted (negative value).  Measuring
the residual ``energies'' is similar to computing a $\chi^2$ in
least-square fitting. Another way to quantify the quality of the fit is to
compare the standard deviation of the residuals
$\sigma_r=\left(\frac{1}{K-1}\sum_n r_{n,l}^2\right)^{1/2}$ with the noise
standard deviation $(\sigma_b)$ on the observed spectrum. The fit is good
when $\sigma_r \sim \sigma_b$.

We can in addition use the CRB framework to filter out pixels with
inaccurate estimations. As explained in Sect.~\ref{sec_accurate_detection},
an estimation will be considered inaccurate when there is at least one
estimation among the $3\times 3$ neighboring pixels, for which at least one
of the following conditions is satisfied
\eq{ %
  \begin{array}{ll}
    {\CRB}^{1/2}(\widehat T_\emr{ex})/\widehat T_\emr{ex}>0.2, & {\CRB}^{1/2}(\log \widehat N)>0.2, \\
    {\CRB}^{1/2}(\widehat\sigma_V) / \widehat\sigma_V >0.2, & {\CRB}^{1/2}(\widehat\Delta_V) / \widehat\sigma_V >0.2.
  \end{array}
  \label{eq_suspicious_2}
}

\FigResidualNoise{}%
\FigResidualEnergyOne{} %

\subsection{Global analysis of the quality of the estimation}
\label{sec_global}

Figure~\ref{Fig_real_data_sigmab_Eb} compares the standard deviation of the
residuals $(\sigma_r)$ with the standard deviation of the noise
$(\sigma_b)$ for all the lines studied here.  If the fits were ideal, the
joint histograms would only peak near a line of slope one.  They thus
suggest that the C$^{18}$O lines are better fitted than the $^{13}$CO
lines, and that the $^{12}$CO lines are the least well fitted. We also
checked that the residuals are larger (i.e., $\sigma_r > \sigma_b$) when
the energy ${\cal E}_x$ is large (not shown in the figure). As the
signal-to-noise ratio also increases with ${\cal E}_x$, this issue implies
some misspecification of the model. The best fit happens for the
C$^{18}$O(1-0) line that has the lowest opacity.  In that case, the
profiles $\Psi$ are almost perfect Gaussian profiles. On the contrary the
high opacity of the $^{12}$CO(1-0) line implies that the profiles are
highly saturated, and any small kinematic perturbations will create
deviations from a Gaussian profile as shown in
Fig.~\ref{Fig_real_data_spectres}, where the asymmetry pattern could not be
taken into account by the model. A better modeling could thus require
additional velocity components especially for the \dcouz line. Another
limit of the model is that it does not encode self-absorption signatures
that may happen at large opacity.

Figures~\ref{Fig_real_data_residus_tcouz}~a and b show the spatial
distributions of the ``energy'' of the observed spectra and of the fit
residuals for the \tcouz line. Both images share the same color look-up
table. On the left image, the yellow pixels correspond to bright molecular
gas while the blue pixels corresponds to faint signal or noise associated
with the IC\,434 H\textsc{ii} region. The ``energy'' of the residual still
exhibits spatially coherent structures, but at a much smaller level than
the ``energy'' of the measured spectra. Our estimator under-fits the
observed spectra as shown in Fig.~\ref{Fig_real_data_residus_tcouz}d. This
may be related to the fact that our model does not fit asymmetric
profiles. However, the fit quality is good as illustrated by the image of
the energy ratio, which shows that the residual ``energy'' amounts to less
than 1\% of the signal (the dark blue color corresponds to 0.01\% in image
\ref{Fig_real_data_residus_tcouz}c) except in regions where the
signal-to-noise ratio becomes small. Figure~\ref{Fig_real_data_residus_all}
shows the same quantities for all the lines modeled in this paper. Overall,
the fitting method is able to recover all the emission with differences at
the percent level or less for all lines.

Increasing the complexity of the line profile model to address the observed
misspecifications could be hazardous. While the increase of the number of
parameters in more complex models certainly allows one to decrease the
difference between the observed spectrum and the model, it may also
increase the variance of estimations in such proportions that obtained
estimations may become useless. In other words, the simple model used here
does not capture all the complexity of the physical processes, but it at
least allows us to capture the processes that it already encodes. The fact
that the residuals are smaller than $1\%$ compared to the observed signal
is sufficient to make the analysis of the excitation temperature, the
column density, and the velocity dispersion pertinent for CO
isotopologues. The main source of systematic errors in the column density
determination results from the deviations from the local thermodynamic
equilibrium, leading to a more complex partition function than the simple
formula in Eq. (9).  The effect is expected to be stronger in warm regions
($T_\emr{ex}\geq 50 - 100\unit{K}$) where many rotational levels are
populated and contribute to the partition function. These warm regions
occupy a small fraction of the total volume and therefore a bias would not
affect the general conclusions. Non-LTE approaches will be developed to
assess more quantitatively the magnitude of the effect and to provide
recommendations on the best method depending on the molecular lines and the
range of physical conditions that are studied.



\newcommand{\TabOne}{%
  \begin{table}
    \centering %
    \caption{Statistics of the estimated parameters over all the pixels for
      the three CO isotopologues.}
    \label{tab_one}
    \resizebox{\hsize}{!}{
      \begin{tabular}{llccc}
        \hline \hline
        Quantity & Unit & $^{12}$CO & $^{13}$CO & C$^{18}$O \\ \hline
        $T_\emr{ex}$ & K & $30 \pm 7.6$ & $17 \pm 4.6$ & $15 \pm 4.4$ \\
        $T_\emr{ex}/T_\emr{dust}$ &  & $1.3 \pm 0.3$ & $0.76 \pm 0.22$ & $0.71 \pm 0.23$ \\
        $\log N $ & cm$^{-2}$ & $18 \pm 0.5$ & $16 \pm 0.45$ & $15 \pm 0.27$ \\
        $\log N/N_{\emr{H}_2}$ &  & $-4.2 \pm 0.4$ & $-5.6 \pm 0.29$ & $-6.7 \pm 0.15$ \\
        $\sigma_V$ & km~s$^{-1}$ & $0.63 \pm 0.17$ & $0.64 \pm 0.26$ & $0.58 \pm 0.2$ \\
        $\log\tau_1$ &  & $0.96 \pm 0.47$ & $-0.06 \pm 0.4$ & $-0.76 \pm 0.25$ \\
        $\log\tau_2$ &  &  & $0.34 \pm 0.41$ & $-0.39 \pm 0.25$ \\ \hline
      \end{tabular}
    }
  \end{table}}

\newcommand{\TabTwo}{%
  \begin{table}
    \centering %
    \caption{Statistics of ratios of the estimated parameters over all the
      pixels for the three CO isotopologues. We stress that
      $\sigma_V(^{12}$CO$)$ is fixed to $\sigma_V(^{13}$CO$)$.}
    \label{tab_two}
    \begin{tabular}{lcc}
      \hline
      \hline
      & $^{13}$CO/C$^{18}$O & $^{12}$CO/$^{13}$CO \\
      \hline
      $T_\emr{ex}/T_\emr{ex}'$ & $1.3 \pm 0.33$ & $1.7 \pm 0.39$ \\
      $\log (N/N') $ & $1.2 \pm 0.17$ & $1.4 \pm 0.33$ \\
      $\sigma_V/\sigma_V'$ & $1.1 \pm 0.25$ & Fixed to $1$ \\
      \hline
    \end{tabular}
  \end{table}}


\newcommand{\FigResults}{%
  \begin{figure*}
    \centering{\includegraphics{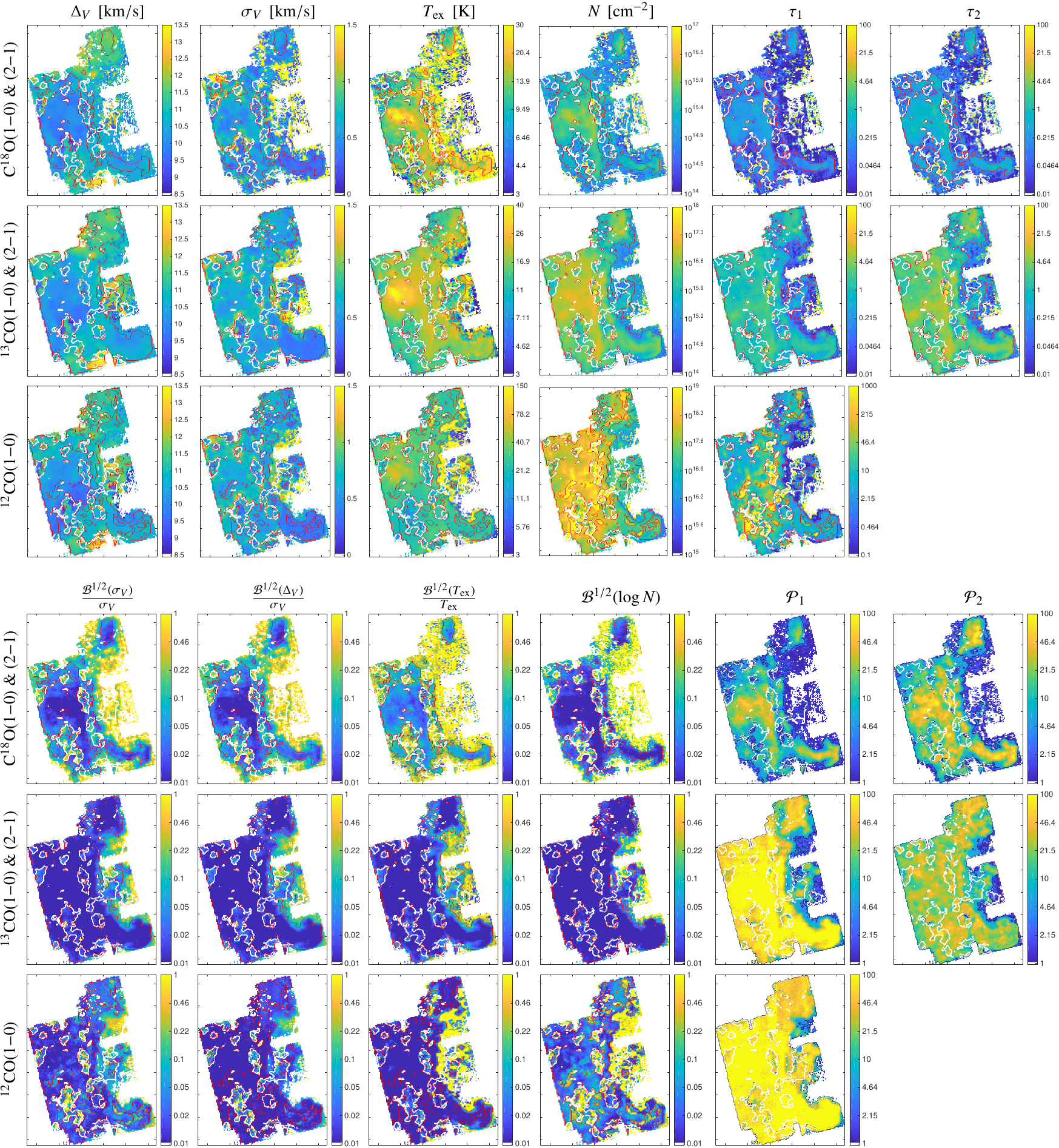}}
    \caption{\textbf{Top:} Spatial variations of the estimated
      physical parameters. \textit{From left to right:} Centroid velocity,
      velocity dispersion, excitation temperature, column density, and line
      opacities.
      \textbf{Bottom:} Spatial variations of the relative
      precisions. \textit{From left to right:} Relative precision on the
      centroid velocity, velocity dispersion, excitation temperature, column
      density, and line peak signal-to-noise ratio.  The black contours on
      the peak signal-to-noise-ratio image delimit the regions where
      ${\cal P} \ge 3$.
      On all images, red contours delimit the regions where the relative
      precision is better than 20\% for all estimated parameters, and white
      contours delimit regions where two components have been estimated. In
      this latter case, the images only show the estimation that is the
      closest (in terms of centroid velocity $\Delta_V$) to its neighboring
      pixels.}
    \label{Fig_real_data_results}
  \end{figure*}
}

\newcommand{\FigTexRatio}{%
  \begin{figure*}
    \centering{\includegraphics{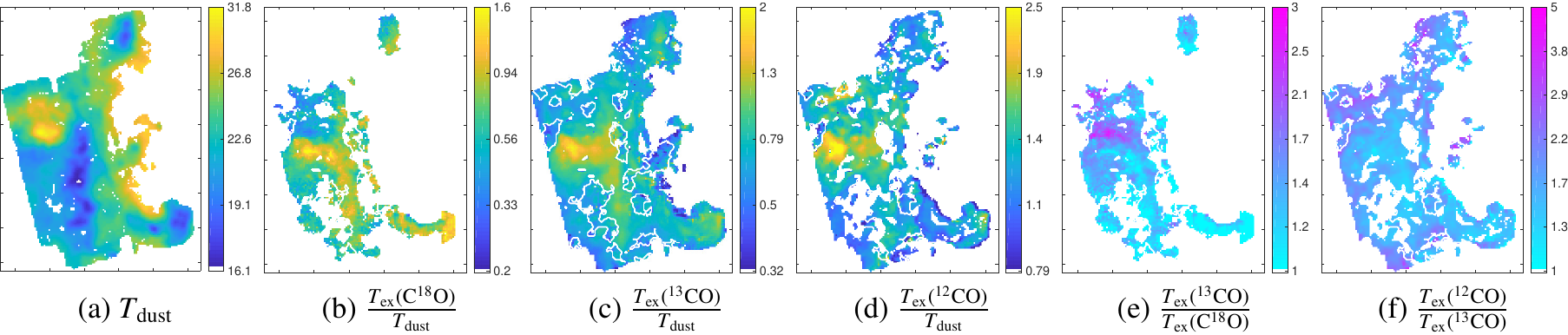}}
    \caption{Comparison between the map of the dust temperature and
      maps of temperature ratios.  Inaccurate estimations are filtered
      out. The dust temperature is only presented in regions with an accurate
      estimation of parameters.}
    \label{Fig_real_data_ima_rapports_T}
  \end{figure*}
}

\newcommand{\FigTexCorrelation}{%
  \begin{figure*}
    \centering{\includegraphics{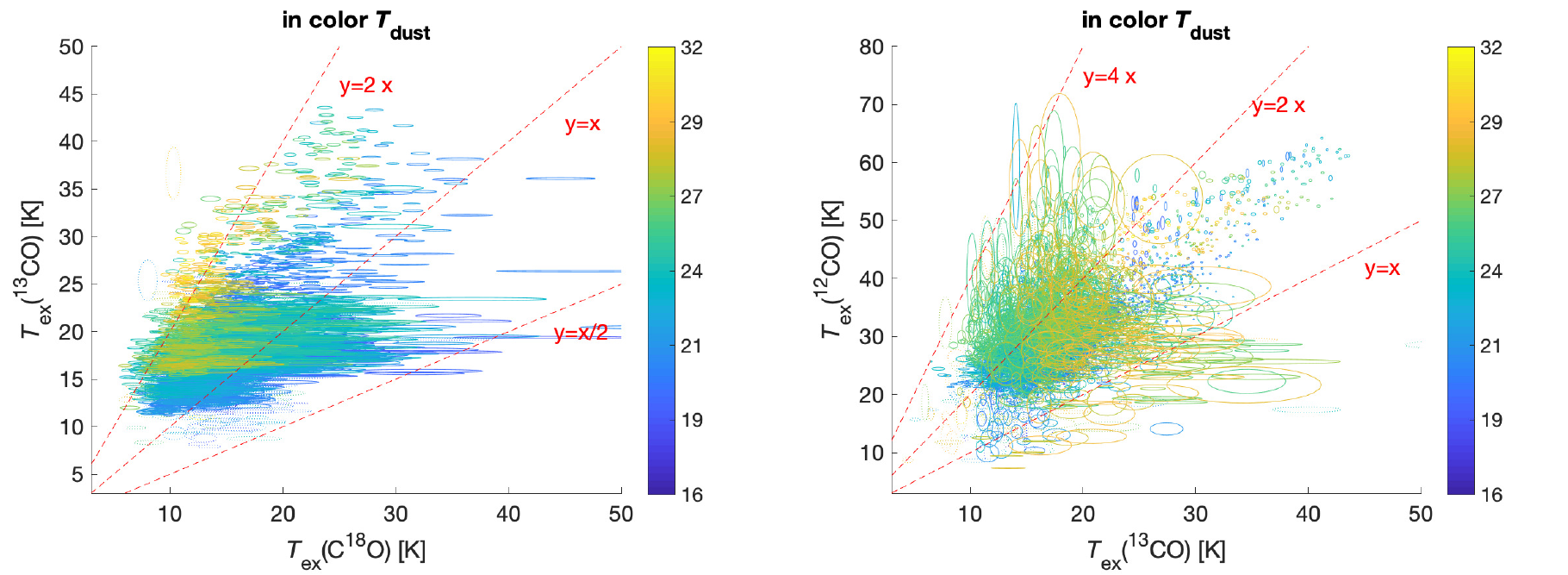}}
    \caption{Scatter plots between the CO isotopologue excitation
      temperatures. The color scale encodes the dust temperature
      $T_\emr{dust}$. The ellipses represent the interval of confidence for
      each estimation: Each ellipse is centered on the estimation of the
      excitation temperature for one pixel and its horizontal and vertical
      sizes are equal to the associated CRBs. Dashed ellipses correspond to
      pixels with two components.  Inaccurate estimations are filtered
      out. Dashed red lines show the loci of ratios 1/2, 1, 2, and 4.}
    \label{Fig_real_data_ellipse_T}
  \end{figure*}
}

\newcommand{\FigEmissivity}{%
  \begin{figure}
    \centering{\includegraphics{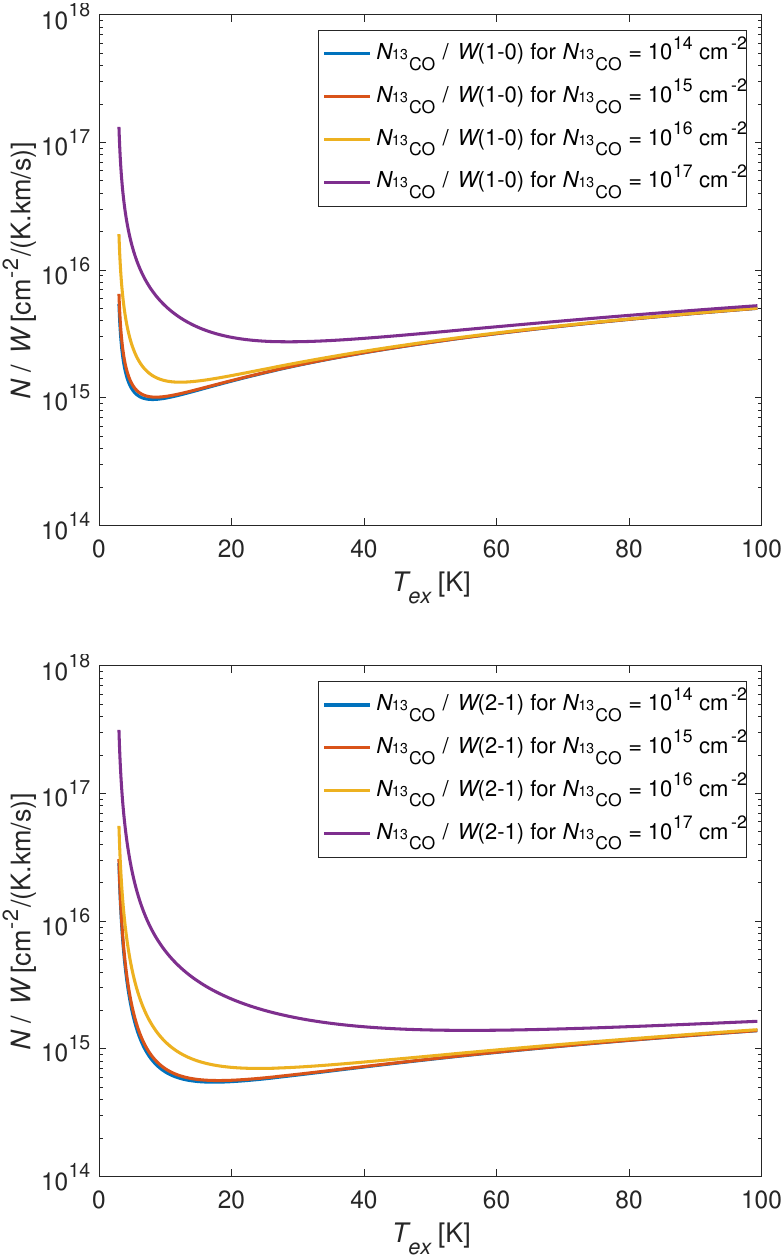}}
    \caption{Plots of the \tco{} column densities per unit intensity of
      \tcouz (\textbf{top}) and \tcodu (\textbf{bottom}) as a function of
      the excitation temperature. This plot is done for four values of the
      \tco{} column density, and $\sigma_V = 0.61\unit{km\,s^{-1}}$.}
    \label{Fig_emissivity}
  \end{figure}
}

\newcommand{\FigColumnRatios}{%
  \begin{figure*}
    \centering{\includegraphics{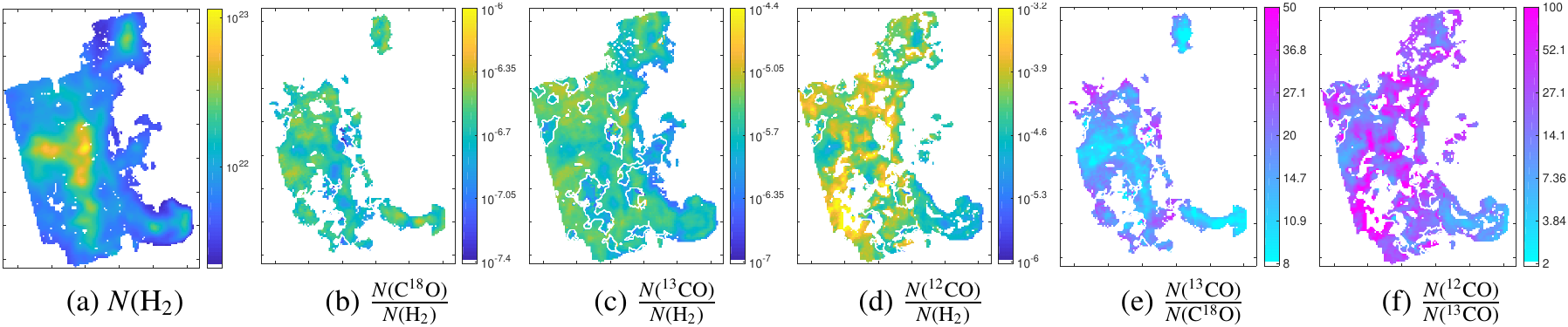}}
    \caption{Comparison of the map of dust-traced H$_2$ column density
      with maps of various column density ratios. Inaccurate estimations were
      filtered out. The dust-traced column density is only presented in
      regions with an accurate estimation of parameters.}
    \label{Fig_real_data_ima_rapports_N}
  \end{figure*}
}

\newcommand{\FigColumnScatter}{%
  \begin{figure*}
    \centering{\includegraphics{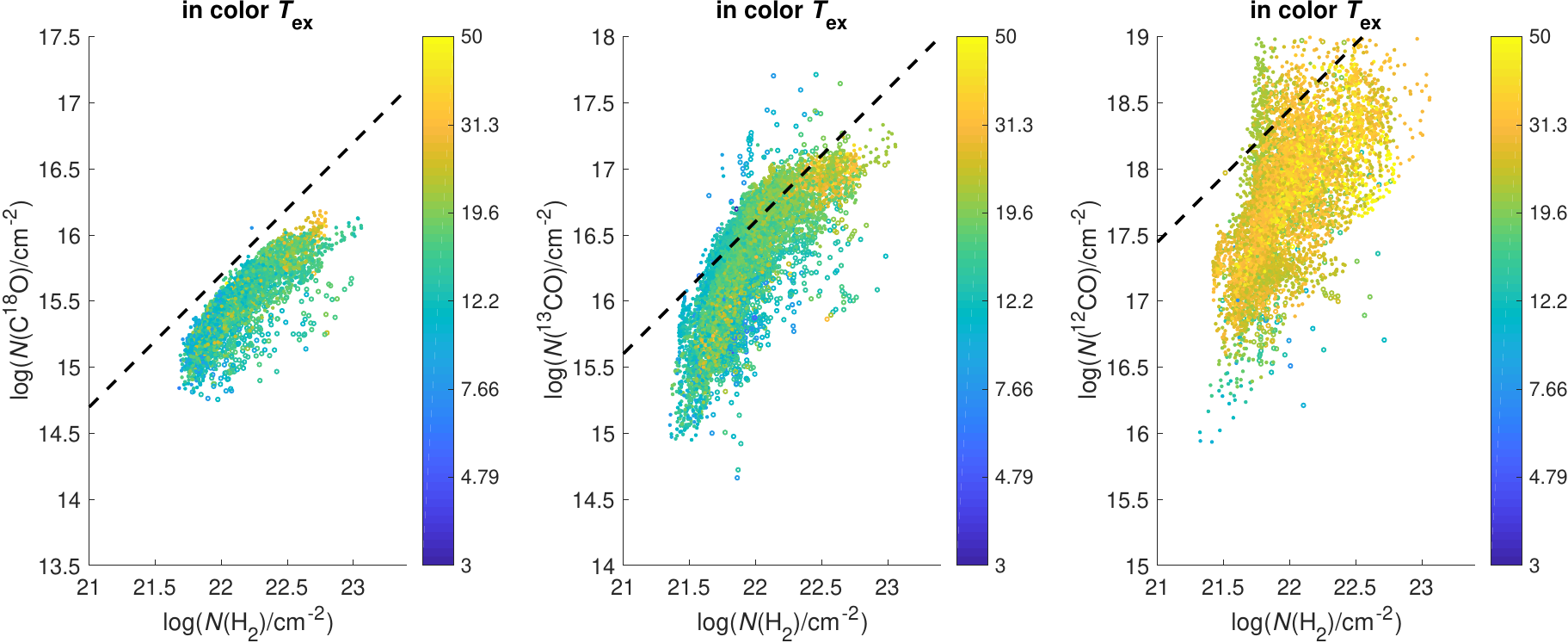}}
    \caption{Scatter plot of the CO isotopologue column densities as a
      function of the dust-traced H$_2$ column density.  Inaccurate
      estimations are filtered out.
      The black dashed lines show the expected gas phase abundances relative
      to H$_2$ with no depletion: C$^{18}$O/H$_2$ $= 5\times 10^{-7}$,
      $^{13}$CO/H$_2$ $= 4\times 10^{-6}$, and CO/H$_2$$ = 1.4 \times
      10^{-4}$.}
    \label{Fig_real_data_scatter_N}
  \end{figure*}
}

\newcommand{\FigColumnCorrelation}{%
  \begin{figure*}
    \centering{\includegraphics{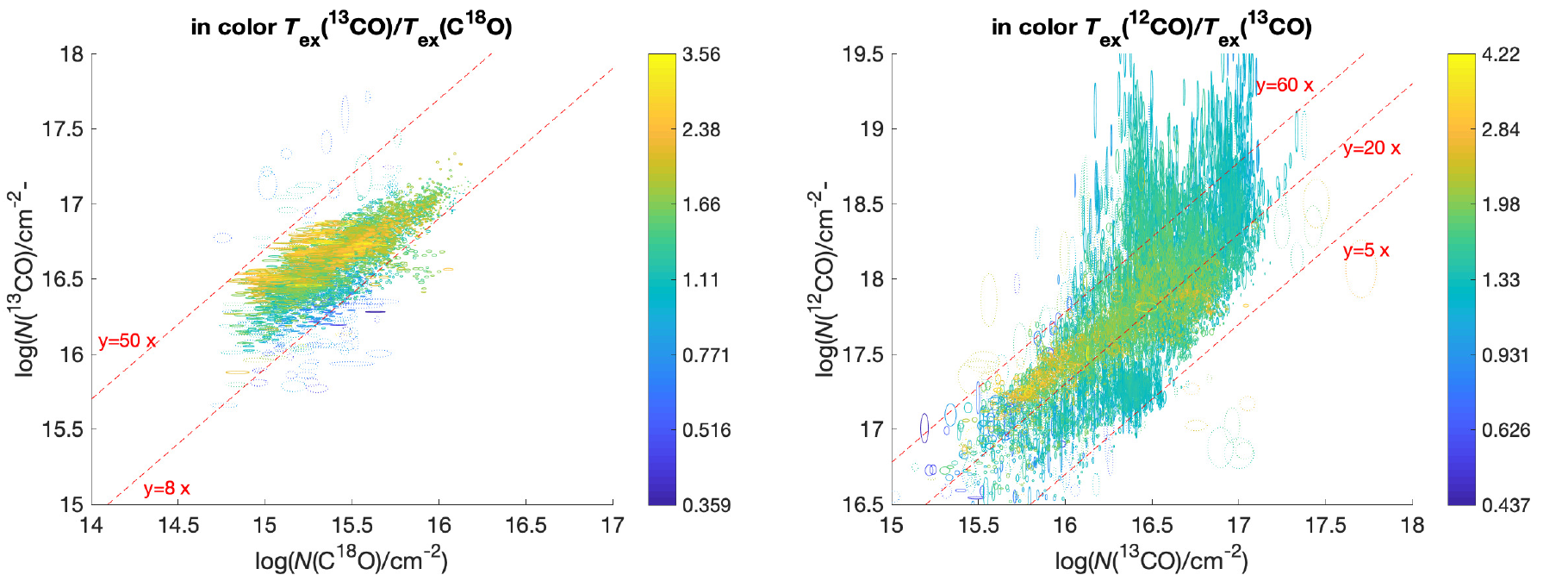}}
    \caption{Scatter plots between CO isotopologue column densities.
      The color scale encodes the ratio of the excitation temperatures of the
      considered species in each panel. The ellipses represent the interval
      of confidence for each estimation: Each ellipse is centered on the
      estimation of the column density for one pixel and its horizontal and
      vertical sizes are equal to the associated CRBs. Dashed ellipses
      correspond to pixels with two components.  Inaccurate estimations are
      filtered out. The dashed red lines show the loci of ratios 5, 20, and
      60.}
    \label{Fig_real_data_ellipse_N}
  \end{figure*}
}

\newcommand{\FigIntensityVsColum}{%
  \begin{figure}
    \centering{\includegraphics{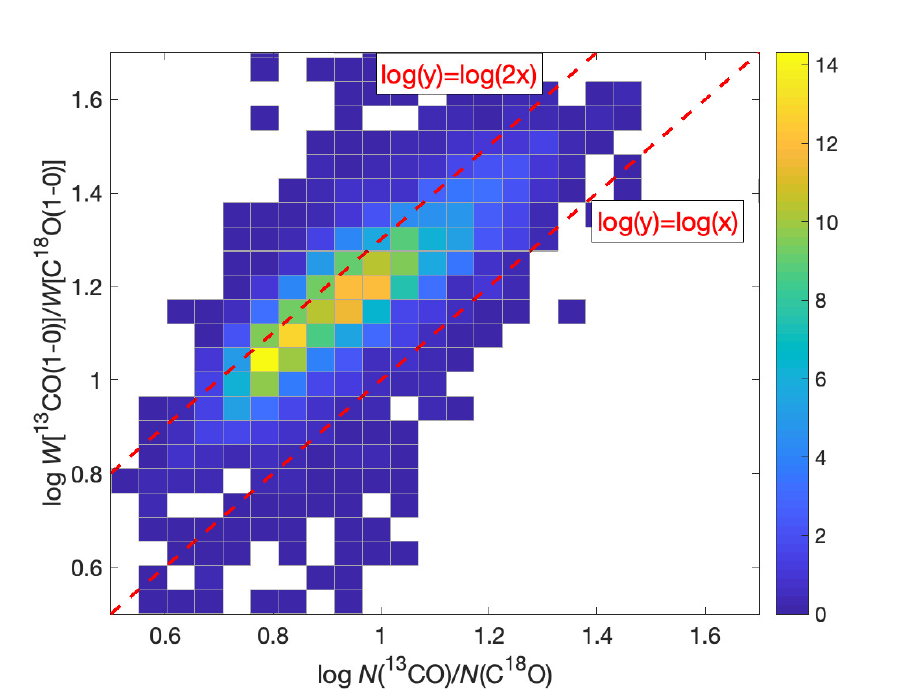}}
    \caption{Joint histogram of the logarithm of the estimated column
      density ratios and of the observed integrated intensity ratios. All the
      ratios are computed for the \tcouz line over the \cdouz one. The dashed
      red lines correspond respectively to ratios of one and two.}
    \label{Fig_Pierre}
  \end{figure}
}


\section{Astrophysical implications}
\label{sec_astro}

\FigResults{}%

The proposed estimator (see Sect.~\ref{sec_ML}) provides accurate column
densities, excitation temperatures, and velocity dispersions in the
framework of LTE excitation and radiative transfer. This allows us to
carefully analyze the errors introduced by the simpler hypotheses that are
commonly used for deriving CO isotopologues column densities.

\subsection{Estimated parameters and associated uncertainties for
  C$^{18}$O, $^{13}$CO, and $^{12}$CO}

\FigTexRatio{}%
\FigTexCorrelation{}%
\TabOne{}%

Figure~\ref{Fig_real_data_results} shows the spatial variations of the
estimated parameters and associated uncertainties for the C$^{18}$O,
$^{13}$CO, and $^{12}$CO isotopologues. Table~\ref{tab_one} lists the mean
and standard deviation values of the excitation temperatures, column
densities, velocity dispersions and opacities. These values are computed
over the field of view that is observed for all the lines.

The peak-signal-to-noise ratio of the \dcouz, \tcouz, and \tcodu lines is
large $(>20)$ over most of the studied field of view. The regions where the
peak-signal-to-noise ratio of the \cdouz and \cdodu is larger than 20 still
amounts to respectively 15\% and 32\% of the studied field of view. The
CRBs are small for all estimated parameters (inside the red contours that
delimit the regions where the relative precision is better than 20\% for
all estimated parameters) except near the regions of transitions between
one and two velocity components (i.e., near the white contours). Even
though the $^{12}$CO signal-to-noise ratio is much larger than the
C$^{18}$O or the $^{13}$CO one, the $^{12}$CO estimations are more
uncertain (see map of ${\cal B}^{1/2}(\log N)$ for $^{12}$CO in
Figure~\ref{Fig_real_data_results}) because a single transition is
available and the $^{12}$CO opacities are large (ranging from 5 to about
1\,000). However, the higher signal-to-noise ratio for $^{12}$CO helps to
derive the velocity field in regions where $^{13}$CO and C$^{18}$O are not
well detected (diffuse gas).

The largest variations are observed in the column density which varies from
the detection limit near 10$^{15}$~cm$^{-2}$ up to values larger than
10$^{17}$~cm$^{-2}$ for $^{13}$CO. The comparison of the column density
maps for $^{13}$CO and C$^{18}$O suggests that the C$^{18}$O molecules are
confined to the high column density regions and avoid the cloud edges. The
$^{12}$CO isotopologue shows a different behavior with emission extending
over most of the imaged field of view and column densities ranging from
$\sim 10^{16}$ to $\sim 10^{19}$ cm$^{-2}$. These behaviors are related to
the difference of opacity of the isotopologues lines. Both $^{13}$CO lines
have moderate opacities across the mapped region, with somewhat higher
values for \tcodu than for \tcouz. The $^{12}$CO is highly optically thick
almost everywhere.

For all CO isotopologues, the excitation temperature presents coherent
spatial variations with maximum values near the NGC\,2023 star forming
region.  Even though the C$^{18}$O emission is fainter and less extended
than that of $^{13}$CO, both species provide similar kinematics
information. The velocity field is spatially regular with well-resolved
gradients, for instance in the Horsehead nebula at the bottom right of the
map. The velocity dispersion $\sigma_V$ ranges from 0.4 to
0.8\unit{km\,s^{-1}}, with values around 0.3\unit{km\,s^{-1}} in the
Horsehead nebula~\citep[see also][]{hily-blant}, and somewhat narrower
lines for C$^{18}$O than for $^{13}$CO.

\subsection{Excitation temperatures}

\FigEmissivity{}%
\TabTwo{}%

Simplifying assumptions are often made when analyzing the CO rotational
emission. The most usual one is that all isotopologues have the same
excitation temperature. It is based on the similarity of the collisional
cross sections. Because the opacity of a rotational transition scales with
the molecular column density, the differences in abundances translate to
different opacities. The main isotopologue ($^{12}$CO) has optically thick
lines while the rarer isotopologues ($^{13}$CO and C$^{18}$O) have lines
either optically thin or with moderate opacities \citep[see,
e.g.,][]{ripple:13}. Another assumption for high density regions
($n > 10^5$ \, cm$^{-3}$ \citealt{goldsmith:78}) is the full thermalization
of the lowest CO rotational levels at the temperature measured on dust
(i.e., assuming the convergence of the dust and gas temperatures).  Both
hypotheses have been questioned. Recently, in their clustering analysis of
a one square degree map in the Orion B cloud, \citet{bron:18} have shown
that the observed CO isotopologue line intensities and line ratios cannot
be explained using these simple hypotheses and that differences in
excitation temperatures should be present.

As shown in Fig.~\ref{Fig_real_data_ima_rapports_T}, the excitation
temperatures of the three isotopologues are different across the field of
view. The $^{12}$CO isotopologue has the largest excitation temperature,
followed by $^{13}$CO and C$^{18}$O. Table~\ref{tab_two} lists the typical
ratios of excitation temperatures. They are
$T_\emr{ex}(^{12}$CO$)/T_\emr{ex}(^{13}$CO$) \sim 1.7$ and
$T_\emr{ex}(^{13}$CO$)/T_\emr{ex}($C$^{18}$O$) \sim
1.3$. Figure~\ref{Fig_real_data_ima_rapports_T} shows that these ratios
vary significantly as a function of the total column density and dust
temperature.
Figure~\ref{Fig_real_data_ellipse_T} suggests that higher dust temperature
regions that trace higher UV illumination conditions, tend to show larger
differences in excitation temperatures between the CO isotopologues. The
same trend is seen when the CO isotopologue excitation temperatures are
compared to the dust temperature as in
Fig~\ref{Fig_real_data_ima_rapports_T}.  A similar effect has been reported
by \cite{Welty:18} for the diffuse/translucent cloud along the line of
sight to HD62542.

The excitation temperature of C$^{18}$O, which traces the UV shielded
regions, is on average lower than $T_\emr{dust}$ (the mean value of
$T_\emr{ex}/T_\emr{dust}$ is 0.71, see Table~\ref{tab_one}). We find a
similar situation for $^{13}$CO(1-0) (mean value
$T_\emr{ex}/T_\emr{dust} = 0.76$), while most of the positions show
$^{12}$CO(1-0) excitation temperatures larger than $T_\emr{dust}$.  This
difference between the CO excitation temperature -- which is a lower
approximation of the gas kinetic temperature-- and the dust temperature is
indeed expected in photo-dissociation regions and UV-dominated regions
where the gas kinetic temperature is larger than the dust temperature. This
is different from the usual approximation that $T_\emr{dust}$ is a good
approximation of the gas kinetic temperature in cold and shielded regions.

The difference in excitation temperatures between CO isotopologues can be
explained by radiative trapping in the $^{12}$CO lines or by the presence
of kinetic temperature gradients along the line of sight, especially near
photodissociation regions, possibly combined with density gradients. Clear
spatial patterns emerge in Fig.~\ref{Fig_real_data_ima_rapports_T}. The
C$^{18}$O and $^{13}$CO excitation temperatures get closer to
$T_\emr{dust}$ in regions where $T_\emr{dust}$ is lower than 20\,K. Because
the dust emission strongly varies with its temperature (as $T^{(4+\beta)}$
where $\beta$ is the dust emissivity index and takes values between 1.5 and
2, \citealt{planck:14}), the dust temperature derived from a single
temperature fit of the spectral energy distribution in lines of sight
combining a strongly UV illuminated region and a more shielded material
does not represent the conditions in the UV shielded region well. The dust
temperature can overestimate the temperature in the shielded gas that
represents the bulk of the material. Somehow the illuminated region
"overshines" as compared to the bulk of the matter.

The error introduced in the column density determination by using an
incorrect excitation temperature can be estimated by examining the
variation of the line column-to-intensity ratio, defined as the ratio of
the column density of the species to the line integrated emission, as a
function of the excitation temperature. Figure~\ref{Fig_emissivity} shows
the variation of the column-to-intensity ratio for the \tcouz{} and
\tcodu{} lines for different column densities of \tco, assuming a velocity
dispersion of $0.61\unit{km\,s^{-1}}$. Column densities of
$N_\emr{^{13}CO} \sim 10^{14}-10^{15}\unit{cm^{-2}}$ correspond to
optically thin lines, while the opacity becomes significant (i.e.,
$\tau\geq 0.5$) for $10^{16}-10^{17}\unit{cm^{-2}}$. In the optically thin
case, the column-to-intensity ratio presents a shallow minimum which
depends on the transition, rises rapidly at temperatures lower than the
minimum and more slowly for excitation temperatures above the minimum.
When the line opacity becomes significant, the shape of the
column-to-intensity ratio curve changes and the minimum shifts to higher
excitation temperatures or possibly disappears.  Therefore, using the
$^{12}$CO excitation temperature for determining the column densities of
$^{13}$CO and C$^{18}$O leads to errors in the estimation of these column
densities because the associated column-to-intensity ratio is
inappropriate. For moderate column densities ($N_\emr{^{12}CO} < 10^{16}$
cm$^{-2}$), the column densities can be underestimated in the case of low
excitation temperatures, or overestimated when using a too high excitation
temperature, depending on the position of $T_\emr{ex}$ relative to the
minimum of the column-to-intensity curve. A CO excitation temperature near
the minimum of the column-to-intensity curve will minimize the error
because a small difference in $T_\emr{ex}$ in this region will not change
the column-to-intensity.  For high column densities
($N \sim 10^{17}\unit{cm^{-2}}$), the bias is significant at low excitation
temperatures ($T_\emr{ex} < 20\,K$) because the column-to-intensity ratio
(purple curve in Fig.~\ref{Fig_emissivity}) is rising fast at low
excitation temperatures.

\subsection{Abundances}

\FigColumnRatios{}%
\FigColumnScatter{}%

Figure~\ref{Fig_real_data_ima_rapports_N} presents maps of the CO
isotopologue abundances relative to H$_2$ (see Sect.~\ref{sec_herschel}),
and Fig.~\ref{Fig_real_data_scatter_N} shows scatter plots of the
relationships between the CO isotopologue column densities and that of
molecular hydrogen.  In the mapped area, $^{13}$CO and $^{12}$CO can be
fitted and analyzed for H$_2$ column densities larger than
$10^{21.5}\unit{cm^{-2}}$. The threshold for C$^{18}$O is about twice
higher at $\sim 10^{21.8}\unit{cm^{-2}}$. Indeed, as shown by
\citet{pety:17} and \citet{orkisz:19}, the threshold for the detection of
C$^{18}$O is $A_V \sim 3\unit{mag}$ or N(H$_2$) = $10^{21.5}\unit{cm^{-2}}$
in the Orion B molecular cloud, while the thresholds for $^{12}$CO and
C$^{13}$O are close to $A_V = 1$\,mag.

Over the mapped area, the mean abundances are well defined at
$N(^{13}$CO$)/N($H$_2)$
$= 10^{-5.6 \pm 0.29} = 2.5 \pm 1.5 \times 10^{-6}$, and
$N($C$^{18}$O$)/N($H$_2)$ $=10^{-6.7\pm0.15} = 2.0 \pm 0.8 \times 10^{-7}$
(see Table~\ref{tab_one}).  These mean abundances are comparable to those
of other molecular clouds in the solar neighborhood such as Taurus and
Ophiuchus \citep{frerking:89}. But these values are about a factor of two
lower than those predicted when one assumes no isotopic fractionation, and
uses the non depleted gas phase carbon elemental abundances applicable to
the Orion region, $\emr{C/H} = 1.4 \times 10^{-4}$, $^{12}$C/$^{13}$C
$= 57 - 67$ and $^{16}$O/$^{18}$O $= 500 - 560$
\citep{gerin:15,langer:90,wilson:94}, namely
$N(^{13}$CO$)/N($H$_2) = 4 - 5 \times 10^{-6}$, and
$N($C$^{18}$O$)/N($H$_2)= 5 - 6 \times 10^{-7}$. The difference is more
pronounced for C$^{18}$O because its abundance is affected by both
photodissociation and freeze-out over a more significant fraction of the
studied area.

Significant deviations from the mean values are present. The C$^{18}$O
abundance is not only lower near the photo-illuminated edges where
molecules are photodissociated, but also in high column density and well
shielded regions. In these latter regions, the dust temperature gets below
the CO condensation temperature ($T_\emr{dust} < 17 $~K), and the CO
molecules can rapidly freeze onto dust grains, lowering the gas phase
abundances. This depletion effect is seen for C$^{18}$O and $^{13}$CO
supporting the explanation by a global freeze-out effect.
 
Because the $^{12}$CO lines are saturated over most of the region, and
because a single line has been observed, the determination of its abundance
is more uncertain. Nevertheless, fixing the value of the velocity
dispersion of $^{12}$CO to the one estimated for $^{13}$CO (see
Sect.~\ref{sec_init}) allows us to determine the $^{12}$CO column density
for the pixels that have the least saturated line emission. Although fairly
uncertain, the resulting values of the $^{12}$CO abundance relative to
H$_2$ are close to $6 \times 10^{-5}$, lower than the expected value using
the carbon elemental abundance relative to H,
$1.4 \times 10^{-4}$~\citep{gerin:15}, but similar to values obtained in
Taurus by \citet{pineda:10}.

\FigColumnCorrelation{}%
\FigIntensityVsColum{}%

\subsection{CO isotopologue column density ratios}

Maps of the CO isotopologue column density ratios are shown in
Fig.~\ref{Fig_real_data_ima_rapports_N} and scatter plots are displayed in
Fig.~\ref{Fig_real_data_ellipse_N}.  With no isotopic fractionation and all
carbon locked in CO, the expected CO isotopologue ratios are
$^{12}$CO/$^{13}$CO $= 57 - 67$ and $^{13}$CO/C$^{18}$O $= 7.5 - 9.8$ using
the elemental abundances given in the previous subsection. As shown in
Fig.~\ref{Fig_real_data_ellipse_N}, the lower bound of the ratio of the
$^{13}$CO and C$^{18}$O column densities is indeed in the expected range at
$N(^{13}$CO$)/N($C$^{18}$O$) = 8$. The upper bound is close to 50
indicating that chemical effects play a significant role, by enhancing the
$^{13}$CO abundance (fractionation) and/or destroying C$^{18}$O
(photodissociation).

Figure~\ref{Fig_Pierre} suggests that the column density ratio is well
correlated with the ratio of line intensities, but the column density ratio
is a factor up to 1.75 smaller than the ratio of line intensities because
of the difference in excitation temperatures and of the moderate opacity of
the $^{13}$CO(1-0) line. Ratios of integrated intensities can therefore be
used to estimate the column density ratios, but after checking with
radiative transfer calculations for a possible multiplicative bias and
introducing a correction if needed.

Although the derivations of $^{12}$CO column densities are uncertain, the
column density ratio $N(^{12}$CO$)/N(^{13}$CO$)$ ranges between 10 and 60.
In particular, the low values of the ratio remain even when the opacity of
the $^{12}$CO line becomes small enough to accurately derive the column
density. Such low values are found in diffuse and translucent gas as a
consequence of efficient fractionation in $^{13}$C due to the exchange
reaction between $^{13}$C$^+$ and $^{12}$CO, that enhances the $^{13}$CO
abundance~\citep{liszt:12}.  The physical conditions in the translucent
envelope of Orion B seem to favor fractionation, which is not restricted to
a small subset of the mapped area but it seen over wide areas. As discussed
by \citet{bron:18} the presence of chemical fractionation over the whole
region can be identified by comparing the ratio of integrated intensities
of the CO isotopologues, and looking at the data in the W(\tco)/W(\dco)
versus W(\tco)/W(\cdo) plane.  The existence of this chemical fractionation
implies that using a single value for the abundance ratios of CO
isotopologues can introduce significant errors when attempting to correct
for the CO opacity in computing its column densities as done by
\cite{Barnes:18} for instance.  This will further increase the bias
introduced by using the same excitation temperature for $^{12}$CO and
$^{13}$CO ground state transitions. In addition to possibly biasing the
results, using such simplifying assumptions is also expected to increase
the dispersion and affect the overall determination of the
$X_\emr{CO} = N(\emr{H_2})/W(\emr{CO})$ conversion factor.

\subsection{Velocity dispersions}

Maps of the velocity dispersions for $^{13}$CO, C$^{18}$O, and $^{12}$CO
are shown in Fig.~\ref{Fig_real_data_results}. The mean values are listed
in Table~\ref{tab_one} and the ratios for the different CO isotopologues
are listed in Table~\ref{tab_two}.  Our formalism includes the line
broadening due to opacity~\citep[see, e.g.,][]{phillips:79}, which is very
significant for $^{12}$CO. This implies that the actual velocity dispersion
derived from high opacity lines like those of $^{12}$CO is smaller than the
apparent line width. Because we fixed the $^{12}$CO velocity dispersion to
the value obtained with $^{13}$CO in our estimation, the velocity
dispersions of $^{12}$CO and $^{13}$CO are identical (see
Sect.~\ref{sec_init}). The velocity dispersions of C$^{18}$O and $^{13}$CO
are however fitted independently.  Both species show similar velocity
dispersions but $^{13}$CO has consistently broader line profiles than
C$^{18}$O. The ratio between the velocity dispersions of $^{13}$CO and
C$^{18}$O is 1.1 (see Table~\ref{tab_two}).  The $^{13}$CO emission is
produced by a more extended volume along the line of sight than the
C$^{18}$O emission as shown by the lower threshold in N(H$_2$) where
$^{13}$CO is detected as discussed above. When analyzing the filamentary
structure of the Orion B molecular cloud, \citet{orkisz:19} showed that the
gas velocity dispersion determined from C$^{18}$O reaches a minimum value
in the filament ridges, and is always lower than the velocity dispersion
determined by $^{13}$CO.  The refined analysis presented here, which takes
the opacity broadening effect into account, confirms the presence of
gradients in velocity dispersion across the spatial extent of the cloud and
along the line of sight, which are captured by the difference between
$^{13}$CO and C$^{18}$O. Inspecting the spatial distribution of the
velocity in Fig.~\ref{Fig_real_data_results} suggests that the small excess
of velocity dispersion for $^{13}$CO relative to C$^{18}$O is more
prominent in the regions with relatively low $T_\emr{dust}$. This supports
the hypothesis that this variation of velocity dispersion is tracing the
starting point of the dissipation of turbulence when entering the dense
filamentary skeleton of the molecular cloud.



\section{Conclusion}

This paper presents an analysis of the precision of the estimation of
physical parameters ($\Delta_V, \sigma_V, N, T_\emr{ex}$) when trying to
fit spectra of low $J$ transitions for the most common CO isotopologues,
using the LTE radiative transfer model. This analysis was based on the
Cramer Rao bound (CRB) computation.
We applied this analysis to the region of the Orion B molecular cloud that
contains the Horsehead pillar and the NGC\,2023 and IC\,434 H{\sc ii}
regions with the following astrophysical results.
\begin{itemize}
\item The mean abundances of the CO isotopologues are consistent with
previous determinations in other regions: $\emr{X(^{12}CO)} \sim 6 \times
10^{-5}$, $\emr{X(^{13}CO)} = 2.5 \pm 1.5 \times 10^{-6}$, and
$\emr{X(C^{18}O)} = 2.0 \pm 0.8 \times 10^{-7}$.
\item The excitation temperatures $T_\emr{ex}$ are different among the CO
isotopologues. $^{12}$CO presents the highest $T_\emr{ex}$, followed by
$^{13}$CO and C$^{18}$O. For $^{13}$CO and C$^{18}$O, the excitation
temperatures are lower than the dust temperature on average, while they are
higher for $^{12}$CO.  This systematic variation can be understood as
resulting from gradients of physical conditions along the line of sight
together with the increased effect of radiative trapping for the more
abundant isotopologues.
\item These differences in $T_\emr{ex}$ imply that the ratio of \tcouz and
\cdouz integrated intensities is not a direct measurement of the column
density ratio $N{\rm{(^{13}CO)}}/N\rm{(C^{18}O)}$, with a difference of up
to a factor two. Moreover, this column density ratio exhibits regular
spatial variations across the mapped region, with high values in the UV
illuminated regions and low values in shielded regions.  These low values
are consistent with the ratio of $^{13}$C and $^{18}$O elemental abundances
(i.e., a factor of about 8).
\item In this nearby molecular cloud, the elemental abundances are uniform
and the variations in CO isotopologue relative abundances are solely due to
chemical processes (fractionation, photodissociation, freezing). The
interpretation of variations of line integrated intensity ratios should
therefore be performed with caution, taking into account radiative transfer
and chemical effects.
\end{itemize}

We obtained the following results from the methodological viewpoint.
\begin{itemize}
\item This analysis has shown that it is important to take the opacity
  broadening effect into account when fitting the line profiles, even for
  moderate opacities as first discussed by \cite{phillips:79}. The
  estimation of the column density is correlated with that of the velocity
  dispersion when the line is optically thick and this correlation between
  column density and velocity dispersion must be taken into account when
  estimating uncertainties on the fitted parameters even for moderate line
  opacities. When the line is optically thin, the estimation of the column
  density is correlated with that of the excitation temperature except in a
  small interval where the ratio of the column density of the species to
  the integrated intensity of the line reaches a minimum (around
  $T_{\rm{ex}} = 10$ K for the 1-0 transitions of CO isotopologues). This
  means that a small variation of the estimation of the excitation
  temperature or velocity dispersion will lead to large errors on the
  estimation of the column density.
\item This analysis also allows us to quantify the benefit of a
  simultaneous analysis of two rotational lines of the same species
  compared to the analysis of a single line. In particular, it alleviates
  the degeneracy described above.  This is a rigorous demonstration of
  intuitive results. It is an argument in favor of the installation of
  dual-band receiver systems for telescopes like the IRAM-30m or NOEMA.
\item In order to derive the precision achieved on these parameters when
  trying to fit actual observations of the CO isotopologue lines towards
  the Orion B molecular cloud, we first showed that a (simple) maximum
  likelihood estimator is unbiased and efficient when the relative
  precision given by the CRB is better than $20\%$, and that it is possible
  to detect pixels for which the estimation of the parameters in LTE
  conditions is accurate (i.e., better than $20\%$).
\item The residuals of the fit of the CO isotopologue lines amount to less
  than 1\% of the original signal and the relative precision on the
  physical parameters is better than 20\% for 63\%, 82\%, and 40\% of the
  field of view for the $^{12}$CO, $^{13}$CO, and C$^{18}$O species,
  respectively. The presence of structured residuals nevertheless indicates
  that the model remains sometimes too simple. In particular, asymmetric
  line profiles or the presence of line wings are incorrectly modeled. The
  $^{12}$CO line profiles are the most affected. Addressing the possibility
  of catching the complex shape of this spectrum is a motivating
  perspective that would generalize the approach initiated in this paper.
\end{itemize}

In the transition between regions where the number of required velocity
components changes, some ambiguity on the velocities of the components
occurs, and this impacts the estimations of all the other
parameters. Fixing \textit{a priori} $\Delta_V$ based on spatial processing
(e.g. extending the ROHSA pre-processing) and thus applying only a gradient
on $\sigma_V$, $T_\emr{ex}$ and $N$ could improve the robustness of our
estimations. This would be useful, in particular, for low signal-to-noise
ratio pixels. Another perspective is to use a spatial regularization
criterion in the fit to improve all the estimations. This will be the
subject of another paper (Vono et al., in prep.). Finally, trying to
estimate the above physical parameters in regions that are more diffuse
than on the studied field of view or for other species that have higher
dipole moments (HCO$^+$ or CS), requires the use of non-LTE models. Such
non-LTE models would also be interesting to identify possible systematic
effects coming from the use of the LTE approximation. Another paper will
study the Large Velocity Gradient approximation of the radiative transfer
using a similar CRB approach.


\begin{acknowledgements}
  This work is based on observations carried out under project numbers
  019-13, 022-14, 145-14, 122-15, 018-16, and finally the large program
  number 124-16 with the IRAM 30m telescope. IRAM is supported by INSU/CNRS
  (France), MPG (Germany) and IGN (Spain). This research also used data
  from the Herschel Gould Belt survey (HGBS) project
  (http://gouldbelt-herschel.cea.fr). The HGBS is a Herschel Key Programme
  jointly carried out by SPIRE Specialist Astronomy Group 3 (SAG 3),
  scientists of several institutes in the PACS Consortium (CEA Saclay,
  INAF-IFSI Rome and INAF-Arcetri, KU Leuven, MPIA Heidelberg), and
  scientists of the Herschel Science Center (HSC).
  We thank CIAS for their hospitality during the many workshops devoted to
  the ORION-B project.
  This project has received financial support from the CNRS through
  the MITI interdisciplinary programs.
  This work was supported in part by the Programme National ``Physique et
  Chimie du Milieu Interstellaire'' (PCMI) of CNRS/INSU with INC/INP,
  co-funded by CEA and CNES.
  JRG thanks Spanish MICI for funding support under grant
  AYA2017-85111-P.
  Finally, we thank the anonymous referee for helpful comments on the
  manuscript.
\end{acknowledgements}


\begin{appendix}

\section{Gradient calculation}
\label{sec_Fisher}

The spectrum at frequency $\nu$ is
\eq{
s(\nu)=
\left(J(T_\emr{ex},\nu_l)-J(T_\emr{CMB},\nu)\right)
\left[1-\exp(-\Psi(\nu))\right]
}
It it thus a function of the unknown parameters
$\btheta=[T_\emr{ex},\,\log N,\,\Delta_{V},\,\sigma_{V}]$. The
following gradients are useful to derive the Fisher matrix (see
Eq.~\ref{eq_If}). 

\subsection{$\partial s(\nu)/\partial T_\emr{ex}$}

\eq{
\begin{array}{ll}
\frac{\partial s(\nu)}{\partial T_\emr{ex}}=&
\frac{\partial J(T_\emr{ex}, \nu_l)}{\partial T_\emr{ex}}
\left[1-\exp(-\Psi(\nu))\right] 
\\
&+
\left(J(T_\emr{ex}, \nu_l)-J(T_\emr{CMB},\nu)\right)
\frac{\partial \Psi(\nu)}{\partial T_\emr{ex}}
\exp(-\Psi(\nu))
\end{array}
}

\subsection{$\frac{\partial J(T,\nu)}{\partial T}$}

\eq{
J(T,\nu) = \frac{h \nu}{k}
\frac{1}{\exp{\frac{h \nu}{k T}} - 1}
}
Thus
\eq{
\frac{\partial J(T,\nu)}{\partial T}=
\frac{h \nu}{k}
\frac{
\frac{h \nu}{k T^2}
\exp{\frac{h \nu}{k T}}
}{\left(\exp{\frac{h \nu}{k T}} - 1\right)^2}
=
\frac{h^2 \nu^2}{k^2 T^2}
\frac{\exp{\frac{h \nu}{k T}}}
{\left(\exp{\frac{h \nu}{k T}} - 1\right)^2}
}
and finally
\eq{
\frac{\partial J(T,\nu)}{\partial T}=
\frac{h^2 \nu^2}{k^2 T^2}
\frac{1}
{\exp{\frac{h \nu}{k T}} -2 +\exp{-\frac{h \nu}{k T}}}
}

\subsection{$\frac{\partial \Psi(\nu)}{\partial T_\emr{ex}}$}

\eq{
\Psi(\nu)
=\sum_{l=1}^2
\alpha_l
\phi\left(\nu;\nu_l\left(1-\frac{\Delta_V}{c}\right),
\nu_l \frac{\sigma_V}{c}\right)
\label{eq_Psi_dem}
}
Thus
\eq{
\frac{\partial \Psi(\nu)}{\partial T_\emr{ex}}=
\sum_{l=1}^2
\frac{\partial \alpha_l}{\partial T_\emr{ex}}
\phi\left(\nu;\nu_l\left(1-\frac{\Delta_V}{c}\right),
\nu_l \frac{\sigma_V}{c}\right)
}

\subsection{$\frac{\partial \alpha_l}{\partial T_\emr{ex}}$}

\eq{
\alpha_l= \frac{c^2}{8\pi}\frac{N}{Q(T_\emr{ex})}
\frac{A_l \, g_\emr{up} }{ \nu_l^2} 
\exp\left[-\frac{E_\emr{up}}{T_\emr{ex}}\right]
\left(\exp\left[\frac{h \,\nu_l}{k\,T_\emr{ex}}\right] - 1\right)
\label{eq_alpha_dem}
}
Thus,
\eq{
\frac{\partial \alpha_l}{\partial T_\emr{ex}}=
\left(-\frac{Q'(T_\emr{ex})}{Q(T_\emr{ex})}
+\frac{E_\emr{up}}{T_\emr{ex}^2} 
-\frac{h \,\nu_l}{k\,T_\emr{ex}^2}
\frac{1}{1-\exp -\frac{h \,\nu_l}{k\,T_\emr{ex}}}
\right) \alpha_l
}
where $Q'(T_\emr{ex})=\frac{\partial Q(T_\emr{ex})}{\partial T_\emr{ex}}$ is
numerically computed.

\subsection{$\partial s(\nu)/\partial L_{N}$}

Let us introduce $L_{N}=\log N$.
\eq{
\frac{\partial s(\nu)}{\partial L_{N}}
=
\left(J(T_\emr{ex},\nu_l)-J(T_\emr{CMB},\nu)\right)
\frac{\partial \Psi(\nu)}{\partial L_{N}}
\exp(-\Psi(\nu))
}

\subsection{$\frac{\partial \alpha_l}{\partial L_{N}}$}

Since $N=10^{L_{N}}$ and using Eq.~\eqref{eq_alpha_dem}, one gets 
\eq{
\frac{\partial \alpha_l}{\partial L_{N}}=
\alpha_l\ln(10)
\label{eq_alpha_sur_L}
}
where $\ln$ is the natural logarithm.

\subsection{$\partial \Psi(\nu) /\partial L_{N} $}

From Eq.~\eqref{eq_Psi_dem}, one gets
\eq{
\frac{\partial \Psi(\nu)}{\partial L_{N}}=
\sum_{l=1}^{L}
\frac{\partial \alpha_l}{\partial L_{N}}
\phi\left(\nu;\nu_l\left(1-\frac{\Delta_V}{c}\right),
\nu_l \frac{\sigma_V}{c}\right)
}
Then using Eq.~\eqref{eq_alpha_sur_L}, one gets 
\eq{
\frac{\partial \Psi(\nu)}{\partial L_{N}}
=\log(10) \, \Psi(\nu)
}

\subsection{$\partial s(\nu)/\partial \Delta_{V}$}

\eq{
\frac{\partial s(\nu)}{\partial \Delta_{V}}
=
\left(J(T_\emr{ex}, \nu_l)-J(T_\emr{CMB},\nu)\right)
\frac{\partial \Psi(\nu)}{\partial \Delta_{V}}
\exp(-\Psi(\nu))
}

\subsection{$\partial \Psi(\nu) /\partial \Delta_{V} $}

With $\nu_c=\nu_l\left(1-\frac{\Delta_V}{c}\right)$
and $\sigma_{\nu}=\nu_l \frac{\sigma_V}{c}$
 one has
\eq{
\Psi(\nu)=\sum_{l=1}^2
\alpha_l\,
\phi\left(\nu;\nu_c, \sigma_{\nu}\right)
\label{eq_Psi_dem_2}
}
Thus
\eq{
\frac{\partial \Psi(\nu)}{\partial \Delta_{V}}=
\sum_{l=1}^2
\alpha_l
\frac{\partial \phi\left(\nu; \nu_c, \sigma_{\nu}\right)}
{\partial \Delta_{V}}
}
Since $\phi(\nu;\nu_c,\sigma_\nu)=\frac{1}{\sqrt{2\pi}\sigma_\nu}
\exp\left(-\frac{(\nu-\nu_c)^2}{2\sigma_\nu^2}\right)$, one gets
\eq{
\frac{\partial \Psi(\nu)}{\partial \Delta_{V}}=
-\sum_{l=1}^2
\alpha_l
\frac{\nu_l}{c}
\frac{\partial \phi(\nu;\nu_c, \sigma_{\nu})
}{\partial \nu_c}
}
where
\eq{
\frac{\partial \phi(\nu;\nu_c,\sigma_\nu)}{\partial \nu_c}=\frac{2(\nu-\nu_c)}{2\sigma_\nu^2}\phi(\nu;\nu_c,\sigma_\nu)
}
Thus
\eq{
\frac{\partial \Psi(\nu)}{\partial \Delta_{V}}=
-\sum_{l=1}^2
\alpha_l
\frac{\nu_l}{c}
\frac{(\nu-\nu_c)}{\sigma_\nu^2}
\phi
\left(\nu;\nu_c, \sigma_{\nu}\right)
}

\subsection{$\partial s(\nu) /\partial \sigma_{V} $}

\eq{
\frac{\partial s(\nu)}{\partial \sigma_{V}}
=
\left(J(T_\emr{ex}, \nu_l)-J(T_\emr{CMB},\nu)\right)
\frac{\partial \Psi(\nu)}{\partial \sigma_{V}}
\exp(-\Psi(\nu))
}

\subsection{$\partial \Psi(\nu) /\partial \sigma_{V} $}

From Eq.~\eqref{eq_Psi_dem_2}, one gets
\eq{
\frac{\partial \Psi(\nu)}{\partial \sigma_{V}}=
\sum_{l=1}^2
\alpha_l \,
\frac{\partial \phi\left(\nu; \nu_c, \sigma_{\nu}\right)}
{\partial \sigma_{V}}
}
and with $\phi(\nu;\nu_c,\sigma_\nu)=\frac{1}{\sqrt{2\pi}\sigma_\nu}
\exp\left(-\frac{(\nu-\nu_c)^2}{2\sigma_\nu^2}\right)$, one gets
\eq{
\frac{\partial \Psi(\nu)}{\partial \sigma_{V}}=
\sum_{l=1}^2
\alpha_l
\frac{\nu_l}{c}
\frac{\partial \phi(\nu;\nu_c, \sigma_{\nu})
}{\partial \sigma_\nu}
}
Then,
\eq{
\frac{\partial \phi(\nu;\nu_c,\sigma_\nu)}{\partial \sigma_\nu}=
-\frac{1}{\sigma_\nu}\phi(\nu;\nu_c,\sigma_\nu)
+\frac{2(\nu-\nu_c)^2}{2\sigma_\nu^3}\phi(\nu;\nu_c,\sigma_\nu)
}
Thus
\eq{
\frac{\partial \Psi(\nu)}{\partial \sigma_{V}}=
\sum_{l=1}^2
\alpha_l
\frac{\nu_l}{c}
\left(\frac{(\nu-\nu_c)^2}{\sigma_\nu^3}-\frac{1}{\sigma_\nu}\right)
\phi
\left(\nu;\nu_c, \sigma_{\nu}\right)
}
which can also be written
\eq{
\frac{\partial \Psi(\nu)}{\partial \sigma_{V}}=
\sum_{l=1}^2
\alpha_l
\frac{1}{\sigma_V}
\left(\frac{(\nu-\nu_c)^2}{\sigma_\nu^2}-1\right)
\phi
\left(\nu;\nu_c, \sigma_{\nu}\right).
}



\section{Maximum Likelihood Estimator}
\label{sec_MLE_definition}

\subsection{Definition for two lines of the same species, and a single
  velocity component}

We start with the assumption that we will estimate the physical parameters
$(\btheta)$ of the LTE radiative transfer for two lines of the same
species, and a single velocity component. Starting from Eq.~\eqref{eq_x_n},
we note the $2K$ samples $x_{n,l}$ as $\bchi$. The amount of noise for each
line is fixed to $\sigma_{b,1}$, and $\sigma_{b,2}$.
In this case, the maximum likelihood estimator (MLE) is \citep{gar95}
\eq{ %
  \widehat \btheta= \arg \max_{\btheta}\left( \log l(\btheta;\bchi)
  \right).  }
Thus, $\widehat \btheta$ is the argument that maximizes the likelihood for
the observed sample $\bchi$. With two lines, the likelihood can be written
as
\eq{ %
  l(\btheta;\bchi)= \prod_{l=1}^2 \prod_{n=1}^K
  \frac{1}{\sqrt{2\pi}\sigma_{b,l}}
  \exp\left(-\frac{(x_{n,l}-s_{n,l})^2}{2\sigma_{b,l}^2}.  \right) }
And the log-likelihood is
\eq{ %
  L(\btheta;\bchi)=\mathrm{cte} -\sum_{l=1}^2\frac{\sum_{n=1}^K
    (x_{n,l}-s_{n,l})^2}{2\sigma_{b,l}^2}.
  \label{eq_L_theta}
}

\subsection{Initialization of the unknown parameters}
\label{sec_MLE_init}

The log-likelihood function $L(\btheta;\bchi)$ can have many local
maxima. It is thus crucial to initialize the gradient near the global
maximum. As explained in Sect.~\ref{sec_Fisher_matrix}, the vector of
unknown parameters has 4 components
$(\btheta=[T_\emr{ex},\, \log N,\,\Delta_{V},\,\sigma_{V}]^T)$ in the case
of a single velocity component.
A simple initial estimation of the typical velocity along the line of sight
$(\Delta_V)$ is given by the velocity where the spectrum intensity is
maximum. As the three other unknown parameters ($T_\emr{ex}$, $N$, and
$\sigma_V$) are highly correlated (see section~\ref{sec_CRB_N}), we
systematically search in a 3D grid defined as follows.
\begin{itemize}
\item We sample $N$ logarithmically between $10^{12}$ and
  $10^{18}\unit{cm^{-2}}$ with a step of 0.1 (i.e., 61 values).
\item We sample $T_\emr{ex}$ logarithmically between $3\unit{K}$ and
  $100\unit{K}$ with a step of 0.05 (i.e., 31 values).
\item Finally, we first sample $\sigma_V$ with 0.2, 0.3, ..., 0.6, 1.2,...,
  3.8 km$/$s (i.e., 10 values) before refining the search with a step of
  $0.025\unit{km\,s^{-1}}$ for $\sigma_V\leq 0.6\unit{km\,s^{-1}}$, and of
  $0.05\unit{km\,s^{-1}}$ otherwise.
\end{itemize}
These values have been fixed empirically based on simulations for which we
tried to find a tradeoff between accuracy and computation time.

\subsection{Maximization of the likelihood function through a scoring
  algorithm}
\label{sec_scoring}

The likelihood function is maximized here using Fisher's scoring algorithm
\citep{gar95}. It is an iterative algorithm
\eq{ %
  \widehat \btheta^{(i+1)}=\widehat \btheta^{(i)} + p_i \bI_F^{-1}(\widehat
  \btheta^{(i)}) \nabla_{\btheta} (\widehat \btheta^{(i)})
  \label{eq_scoring}
}
where $i$ is the ith iteration, $p_i$ is a constant, $\bI_F(\btheta)$ is
the Fisher matrix (see Sect.~\ref{sec_CRB}) seen as a function of
$\btheta$, and $\nabla_{\btheta} (\btheta)$ is the gradient
\eq{ %
  \forall j=1,...,4 \quad \left[\nabla_{\btheta} (\btheta)\right]_{j} =
  \sum_{l=1}^2 \frac{1}{\sigma_{b,l}^2}\sum_{n=1}^K \frac{\partial
    s_{n,l}}{\partial \theta_j} (x_{n,l}-s_{n,l}), }
where $j$ is the index of the unknown parameter.

In practice, at each iteration $i$, the algorithm tries $p_i=0.1$, $0.4$
and $0.8$, and it makes a quadratic fit to get an estimation of $p_i$ that
minimizes the log-likelihood in the interval $[0.1,\,0.8]$. Moreover, the
inversion of $\bI_F$ is made with the pseudo-inverse when $\bI_F$ becomes
singular (in the sense that the ratio between the largest and the smallest
eigenvalues of $\bI_F$ is larger than $10^{8}$). Finally, the iteration
loop stops when the log-likelihood verifies
$|L(\widehat\btheta^{(i+1)};\bchi)-L(\widehat\btheta^{(i)};\bchi)|<10^{-16}$
or when the number of iterations reaches 1000.

\subsection{Generalization to two velocity components}
\label{sec_MLE_2_peaks}

When two velocity components are needed, the log-likelihood of
Eq.~\eqref{eq_L_theta} becomes
\eq{ %
  L(\btheta_1,\btheta_2;\bchi)=\mathrm{cte} -\sum_{l=1}^2\frac{\sum_{n=1}^K
    (x_{n,l}-s_{n,l}(\btheta_1) -s_{n,l}(\btheta_2))^2}{2\sigma_{b,l}^2}
  \label{eq_L_theta_two_peaks}
}
where $s_{n,l}(\btheta_m)$ is the spectrum corresponding to the component
$m\in\{1,2\}$. The grid used in the initialization step has now six
dimensions:
$T_\emr{ex}^{(1)},N^{(1)},\sigma_V^{(1)},
T_\emr{ex}^{(2)},N^{(2)},\sigma_V^{(2)}$.  One solution is to consider the
total Fisher matrix of size $8\times 8$ (see section
\ref{sec_two_components}), but it can lead to singularities.  We
empirically observe that using iteratively the gradient on each component
separately ($\btheta_1$, and then $\btheta_2$) actually leads to better
results than using a gradient on the enlarged vector
$[\btheta_1,\,\btheta_2]^T$. Such a coordinate descent only requires
inversions of $4\times 4$ Fisher matrices.

\subsection{Computing load and optimization}

From the computational viewpoint, the estimation will be done many times
either because it will be applied to many different lines of sight or
because it will be used in Monte Carlo simulations. It is thus useful to
actually compute in advance a 5D set of
$\left(s_n(\btheta)\right)_{n,T_\emr{ex},N,\sigma_V,\Delta_V}$ per line.
In this 5D set, $T_\emr{ex}$, $N$ and $\sigma_V$ are sampled as described
in Sect.~\ref{sec_MLE_init}. Moreover, we consider 10 different values of
$\Delta_V$ with a step of $0.05\unit{km\,s^{-1}}$ and we restrict the range
of explored channels to the velocity range where the lines appear, i.e., an
interval of 26.5\unit{km\,s^{-1}} around the initially estimated
$\Delta_V$. This last point decreases substantially the computation time.
For a single species, the computation of the 5D set of
$\left(s_n(\btheta)\right)_n$ takes around 13 seconds in our Matlab
implementation on a standard 2016 laptop.  Each subsequent estimation
(initialization plus gradient) takes around 0.05 or 1.0 second when
estimating one or two velocity components, respectively.

For the considered ORION-B data, and with the initialization proposed in
Sect.~\ref{sec_MLE_init}, the median number of iterations required to reach
convergence is 40 or 200 when estimating one or two velocity components,
respectively.



\newcommand{\FigMLEone}{%
  \begin{figure}
    \centering{\includegraphics{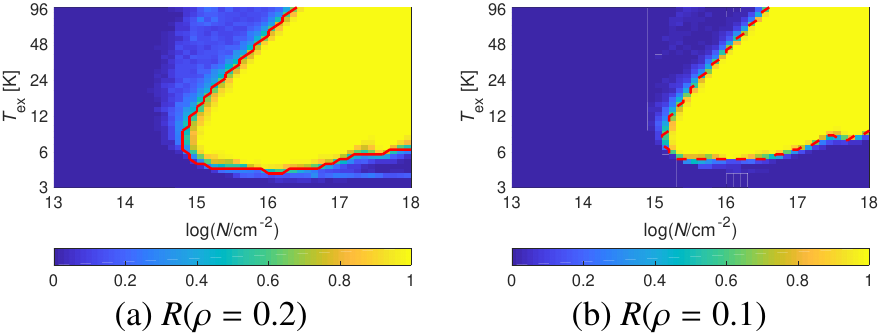}}
    \caption{Variations of the relative number of accurate estimations
      as a function of the column density and excitation temperature.
      The contour corresponds to the frontier where the relative precision
      on the actual values of the four parameters is $\rho=20\%$
      (\textbf{solid red contour on the left}), and $10\%$ (\textbf{dashed
        red contour on the right}).  The \tcouz and \tcodu lines are
      simulated with $\sigma_b=\ValSigmab$\unit{mK},
      $\Delta_V=1.1$\unit{km\,s^{-1}}, and
      $\sigma_V=0.61$\unit{km\,s^{-1}}.}
    \label{Fig_prop}
  \end{figure}
}

\newcommand{\FigMLEtwo}{%
  \begin{figure}
    \centering{\includegraphics{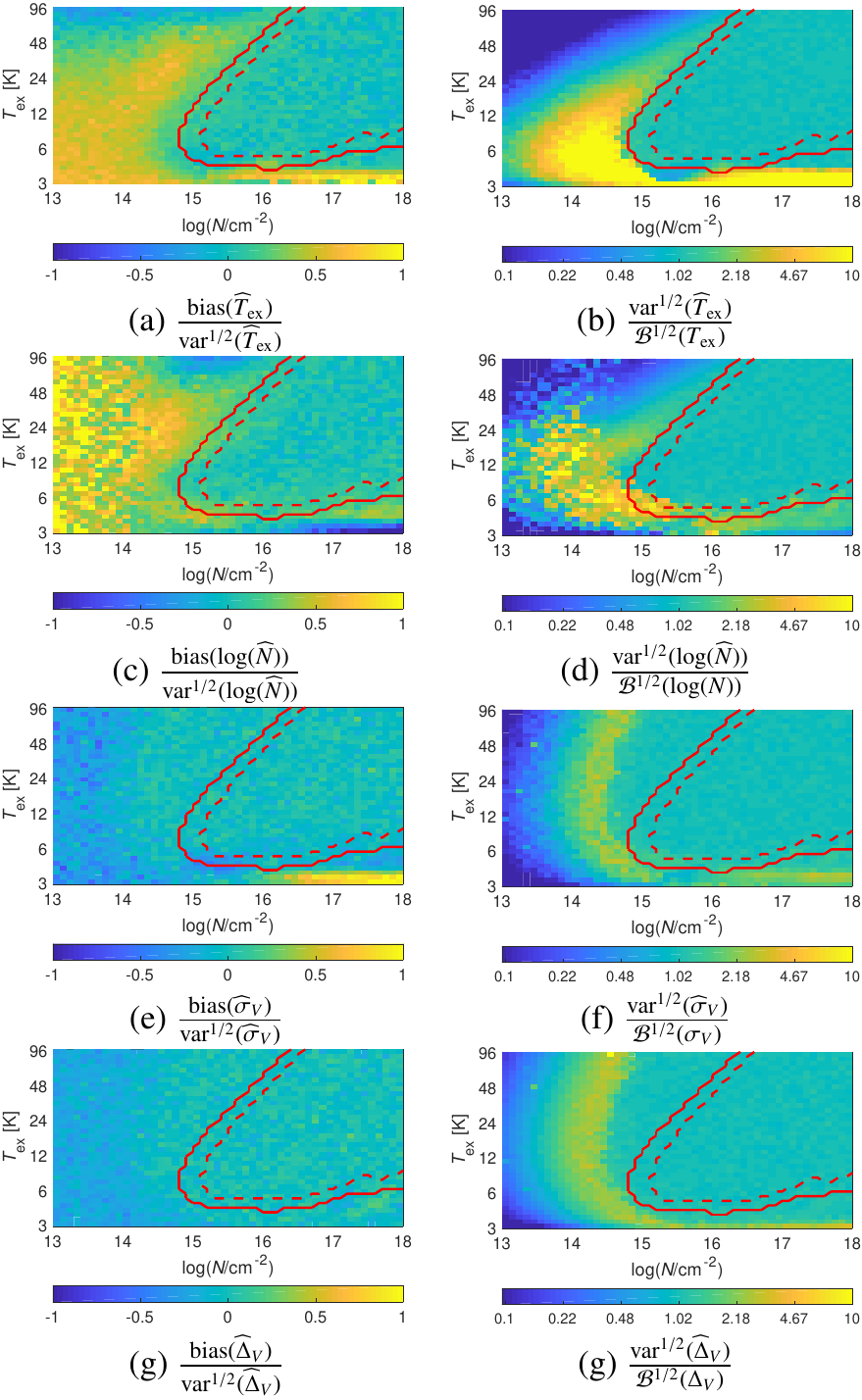}}
    \caption{Variations of the bias (\textbf{left}) and efficiency
      (\textbf{right}) of the maximum likelihood estimator as a function of
      the column density and excitation temperature.
      The solid and dashed red contours correspond to the frontiers where the
      relative precisions on the estimations on the four parameters are
      $\rho=20$, and $10\%$, respectively, as defined in Fig.~\ref{Fig_prop}.
      The \tcouz and \tcodu lines are simulated with
      $\sigma_b=\ValSigmab$\unit{mK}, $\Delta_V=1.1\unit{km\,s^{-1}}$, and
      $\sigma_V=0.61\unit{km\,s^{-1}}$.}
    \label{Fig_var_13co10_13co21_610}
  \end{figure}
}


\section{Performance of the Maximum Likelihood Estimator}
\label{sec_MLE_performance}

\FigMLEone{} %
\FigMLEtwo{} %

Monte Carlo simulations with $P$ independent realizations of the estimator
$\left\{\widehat\btheta^{(p)}\right\}_{p=1,...,P}$ are used to analyze its
performance. We simulate data for the \tcouz and \tcodu lines using the
different values of $T_\emr{ex}$ and $N$ already used in
Sect.~\ref{sec_CRB_varitions}. We here compute $P=200$ simulations with
random noise for each pair of ($T_\emr{ex},N$) values.

After showing that MLE estimates of the parameters $(\widehat\btheta)$ can
be injected in the computation of the CRB to detect accurate estimations,
we will show that the proposed estimator is unbiased and efficient.

\subsection{Detection of (in)accurate estimations}
\label{sec_accurate_detection}

In order to physically interpret the estimations $\widehat\btheta$ computed
on observed data, it is crucial to remove inaccurate estimations. We thus
need a way to quickly detect inaccurate estimations. Figure~\ref{Fig_prop}
shows the fraction of the Monte Carlo realizations for each pair of
($T_\emr{ex},N$) values that deliver ``accurate'' estimations of all
parameters. ``Accurate'' here means that all the following conditions are
simultaneously satisfied
\eq{ %
  \begin{array}{ll}
    {\CRB}^{1/2}(\widehat T_\emr{ex})/\widehat T_\emr{ex} \leq\rho, & {\CRB}^{1/2}(\log \widehat N) \leq\rho, \\
    {\CRB}^{1/2}(\widehat\sigma_V) / \widehat\sigma_V \leq\rho,     & {\CRB}^{1/2}(\widehat\Delta_V) / \widehat\sigma_V \leq\rho.
  \end{array}
  \label{eq_suspicious}
}
In these equations, $\rho$ is the relative precision required for all the
parameters. At first sight, Eq.~\eqref{eq_suspicious} and
Eq.~\eqref{eq_rho} seem identical. However, we here use estimation of the
parameters $\widehat\btheta$, while we used values of $\btheta$ used to
simulate the data in Sect.~\ref{sec_CRB_relative}.  The value $R(\rho)$ is
the fraction of ``accurate'' estimations {\it detected without a priori
  information} on the parameters.

The contours in Fig.~\ref{Fig_prop} correspond to the frontiers where the
relative precision on the actual values $\btheta$ of the four parameters is
$\rho=20\%$ or $10\%$. Figure~\ref{Fig_prop} thus clearly suggests that
these frontiers are close to the pixels for which $R(\rho)\simeq 0.5$. The
test $R(\rho)\geq 0.5$ thus gives a fair detection of accurate (and
inaccurate) estimations.
Furthermore, these maps show that, most of the time, all $P$ estimations
are either inaccurate ($R(\rho)=0$ in blue) or accurate ($R(\rho)=1$, in
yellow).  In other words, a single estimation $\widehat\btheta$ is often
sufficient to detect whether it is accurate or not.

More precisely, the light blue area in Fig.~\ref{Fig_prop}.a corresponds to
a value $R(0.2)=0.15$. This means that we still have a $15\%$ chance of
considering an estimation as accurate when it is in fact inaccurate, when
$N\simeq 10^{15}\unit{cm^{-2}}$ and $T_\emr{ex}>50\unit{K}$. It is possible
to improve this on observed data because adjacent pixels on the sky have
physical parameters that are partially correlated.  We can thus assume that
accurate and inaccurate estimations are spatially grouped, and the
detection can be improved by computing Eq.~\eqref{eq_suspicious} in a
sliding window of size $3\times 3$ pixels. In practice, we remove
estimations of pixels for which one of the neighbors in the $3\times3$
pattern violates Eq.~\eqref{eq_suspicious}.

\subsection{Bias and variance of the estimator}

The bias and variance of the maximum likelihood estimator can be estimated
with \citep{gar95}
\eq{ %
  \begin{array}{c}
    \widehat \bias(\widehat \theta_i)=\frac{1}{P}\sum \widehat \theta^{(p)}_i-\theta_i \\
    \widehat \var(\widehat \theta_i)=\frac{1}{P-1}\sum \left(\theta^{(p)}_i-\frac{1}{P}\sum \widehat \theta^{(p)}_i\right)^2,
  \end{array}
} where $P$ is the number of simulations in the Monte Carlo analysis,
$\theta_i$ are the actual values of the parameters, and $\widehat \theta_i$
are the estimated values.
Figure~\ref{Fig_var_13co10_13co21_610} shows the variations of the ratios
$\widehat \bias(\widehat \theta_i) / \widehat \var^{1/2}(\widehat
\theta_i)$ and
$\widehat \var^{1/2}(\widehat \theta_i) / {\cal B}^{1/2}(\theta)$ as a
function of $T_\emr{ex}$ and $N$.

It shows that the bias of the proposed estimator is negligible compared to
its standard deviation (i.e.,
$\widehat \bias(\widehat \theta_i) \ll \widehat \var^{1/2}(\widehat
\theta_i)$), and that its variance reaches the Cramer Rao bound (i.e.,
$\widehat \var(\widehat \theta_i) \simeq \CRB(\theta_i)$) in the region
where the CRBs are sufficiently small to get accurate estimations.



\newcommand{\FigResidualEnergyAll}{%
  \begin{figure*}
    \centering{ \includegraphics{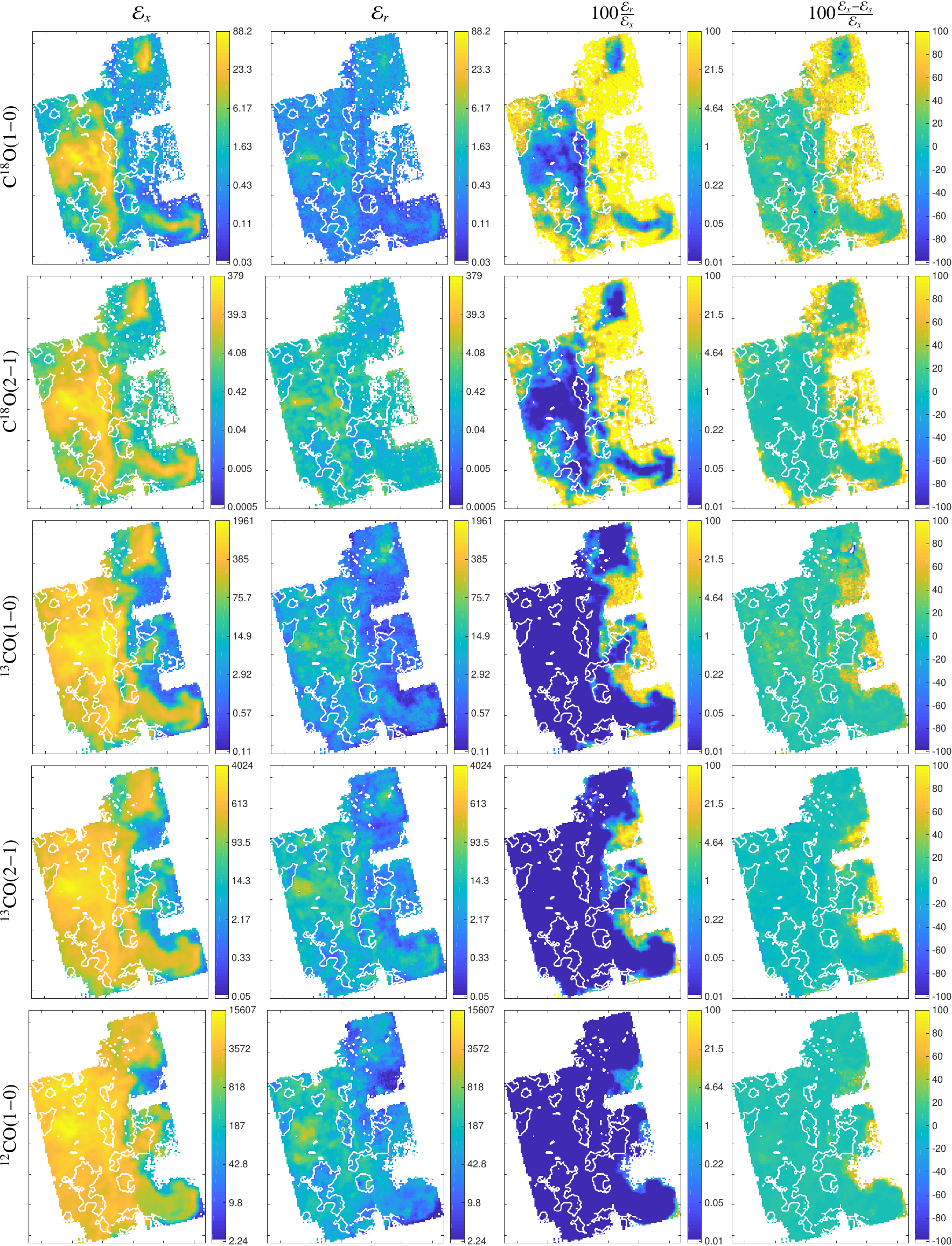}}
    \caption{Spatial variations of the observed spectrum ``energy''
      (\textbf{first column}), of the residual ``energy'' (\textbf{second
        column}), of their ratio (\textbf{third column}), and of the ratio of
      ``energy'' that has not been modeled. The unit of the color look-up
      table is Kelvin$^{2}$ or \% depending on the column.  White contours
      show the regions where two components have been detected.}
    \label{Fig_real_data_residus_all}
  \end{figure*}
}

\section{Additional figures}

In section~\ref{sec_global}, Figure~\ref{Fig_real_data_residus_tcouz} shows
the residuals for only one line. Figure~\ref{Fig_real_data_residus_all} is
a generalization to the other lines.

\FigResidualEnergyAll{} %


\end{appendix}

\end{document}